\documentclass[aps,prd,nofootinbib,floatfix]{revtex4}
\usepackage{graphicx}
\usepackage{rotating}

\begin{document}
\title{Multiple inflation and the {\sl WMAP} `glitches'\\
       II. Data analysis and cosmological parameter extraction} 
\author{Paul Hunt$^{1,2}$ and Subir Sarkar$^{1}$}
\medskip
\affiliation{$^{1}$ Rudolf Peierls Centre for Theoretical Physics, University
             of Oxford, 1 Keble Road, Oxford OX1 3NP, UK \\
             $^{2}$ Institute of Theoretical Physics, Warsaw
             University, ul Ho\.za 69, 00-681 Warsaw, POLAND}

\begin{abstract}
Detailed analyses of the {\sl WMAP} data indicate possible oscillatory
features in the primordial curvature perturbation, which moreover
appears to be suppressed beyond the present Hubble radius. Such
deviations from the usual inflationary expectation of an approximately
Harrison-Zeldovich spectrum are expected in the supergravity-based
`multiple inflation' model wherein phase transitions during inflation
induce sudden changes in the mass of the inflaton, thus interrupting
its slow-roll. In a previous paper we calculated the resulting
curvature perturbation and showed how the oscillations arise. Here we
perform a Markov Chain Monte Carlo fitting exercise using the 3-year
{\sl WMAP} data to determine how the fitted cosmological parameters
vary when such a primordial spectrum is used as an input, rather than
the usually assumed power-law spectrum. The `concordance' $\Lambda$CDM
model is still a good fit when there is just a `step' in the
spectrum. However if there is a `bump' in the spectrum (due e.g. to
two phase transitions in rapid succession), the precision CMB data can
be well-fitted by a flat Einstein-de Sitter cosmology {\em without}
dark energy. This however requires the Hubble constant to be $h \simeq
0.44$ which is lower than the locally measured value. To fit the {\sl
SDSS} data on the power spectrum of galaxy clustering requires a $\sim
10\%$ component of hot dark matter, as would naturally be provided by
3 species of neutrinos of mass $\sim 0.5$ eV. This CHDM model cannot
however fit the position of the baryon acoustic peak in the LRG
redshift two-point correlation function. It may be possible to
overcome these difficulties in an inhomogeneous
Lema\^itre-Tolman-Bondi cosmological model with a local void, which
can potentially also account for the SN~Ia Hubble diagram without
invoking cosmic acceleration.
\end{abstract}

\maketitle

\section{Introduction}

Precision measurements of CMB anisotropies by the {\sl Wilkinson
  Microwave Anisotropy Probe} ({\sl WMAP}) are widely accepted to have
firmly established the `concordance' $\Lambda$CDM model --- a flat
universe with $\Omega_\Lambda \simeq 0.7$, $\Omega_\mathrm{m} \simeq
0.3$, $h \simeq 0.7$, seeded by a nearly scale-invariant power-law
spectrum of adiabatic density fluctuations
\cite{Spergel:2003cb,Spergel:2006hy}. However the model fit to the
data is surprisingly poor. For the {\sl WMAP}-1 TT spectrum,
$\chi_{\rm eff}^2/\nu = 974/893$ \cite{Spergel:2003cb}, so formally
the $\Lambda$CDM model is ruled out at $97\%$ c.l. Visually the most
striking discrepancies are at low multipoles where the lack of power
in the Sachs-Wolfe plateau and the {\em absence} of the expected
$\Lambda$-induced late integrated-Sachs-Wolfe effect have drawn much
attention. However because cosmic variance and uncertainties in the
foreground subtraction are high on such scales, it has been argued
that the observed low quadrupole (and octupole) are not particularly
unlikely, see e.g.
refs.\cite{Efstathiou:2003wr,Slosar:2004xj,Park:2006dv}. The excess
$\chi^2$ in fact originates mainly from sharp features or `glitches'
in the power spectrum that the model is unable to fit
\cite{Spergel:2003cb,Hinshaw:2006ia}. Although these glitches are less
pronounced in the 3-year data release, they are still present
\cite{Hinshaw:2006ia}. Hence although the fit to the concordance
$\Lambda$CDM model has improved with $\chi_{\rm eff}^2/\nu = 1049/982$
for the {\sl WMAP}-3 TT spectrum \cite{Spergel:2006hy}, this model
still has only a 6.8\% chance of being a correct description of the
data. This is less than reassuring given the significance of such a
tiny cosmological constant (more generally, `dark energy') for both
cosmology and fundamental physics.

The {\sl WMAP} team state: {\it ``In the absence of an established
  theoretical framework in which to interpret these glitches (beyond
  the Gaussian, random phase paradigm), they will likely remain
  curiosities''} \cite{Hinshaw:2006ia}. However it had been noted
earlier by the {\sl WMAP} team themselves \cite{Peiris:2003ff} that
these may correspond to sharp features in the spectrum of the
underlying primordial curvature perturbation, arising due to sudden
changes in the mass of the inflaton in the `multiple inflation' model
proposed in ref.\cite{Adams:1997de}. This generates characteristic
localized oscillations in the spectrum, as was demonstrated
numerically in a toy model of a `chaotic' inflationary potential
having a `step' starting at $\phi_\mathrm{step}$ with amplitude and
gradient determined by the parameters $c_\mathrm{ampl}$ and
$d_\mathrm{grad}$: $V(\phi)=\frac{1}{2}m_{\phi}^2\phi^2 \left[1 +
  c_\mathrm{ampl}\tanh
  \left(\frac{\phi-\phi_\mathrm{step}}{d_\mathrm{grad}}\right)\right]$
\cite{Adams:2001vc}. It was found that the fit to the {\sl WMAP}-1
data improves significantly (by $\Delta\chi^2 = 10$) for the model
parameters $\phi_\mathrm{step} = 15.5\,M_\mathrm{P}$, $c_\mathrm{ampl}
= 9.1\times10^{-4}$ and $d_\mathrm{grad} =
1.4\times10^{-2}\,M_\mathrm{P}$, where
$M_\mathrm{P}\equiv(8\pi\,G_\mathrm{N})^{-1/2} \simeq
2.44\times10^{18}$~GeV \cite{Peiris:2003ff}. However the cosmological
model parameters were held fixed (at their concordance model values)
in this exercise. Recently this analysis has been revisited using the
{\sl WMAP}-3 data \cite{Covi:2006ci}; these authors also consider
departures from the concordance model and conclude that there are
virtually no degeneracies of cosmological parameters with the
modelling of the spectral feature \cite{Hamann:2007pa}.

However $m$ in the toy model above is {\em not} the mass of the
inflaton --- in fact in all such chaotic inflation models with $V
\propto \phi^n$ where inflation occurs at field values
$\phi_\mathrm{infl} > M_\mathrm{P}$, the leading term in a Taylor
expansion of the potential around $\phi_\mathrm{infl}$ is always {\em
  linear} since this is not a point of symmetry \cite{German:2001tz}.
The effect of a change in the inflaton mass in multiple inflation can
be sensibly modelled only in a `new' inflation model where inflation
occurs at field values $\phi << M_{\rm Pl}$ and an effective field
theory description of the inflaton potential is possible. The
`slow-roll' conditions are violated when the inflaton mass changes
suddenly due to its (gravitational) coupling to `flat direction'
fields which undergo thermal phase transitions as the universe cools
during inflation \cite{Adams:1997de}. The resulting effect on the
spectrum of the curvature perturbation was found by analytic solution
of the governing equations to correspond to an upward step followed by
rapidly damped oscillations \cite{Hunt:2004vt}.\footnote{A `hybrid'
  inflation model wherein the inflaton is coupled to a `curvaton'
  field also yields oscillations together with suppressed power on
  large scales \cite{Langlois:2004px}. A similar phenomenon had been
  noted earlier for the case where the inflaton potential has a jump
  in its {\em slope} \cite{Starobinsky:ts}; however such a
  discontinuity has no physical interpretation. The {\sl WMAP}
  glitches have also been interpreted as due to the effects of
  `trans-Planckian' physics
  \cite{Martin:2003sg,Danielsson:2006gg,Spergel:2006hy} and due to
  resonant particle production \cite{Mathews:2004vu}.}

One can ask if spectral features are seen when one attempts to recover
the primordial perturbation spectrum directly from the data. Such
attempts may be divided into two classes. Usually the curvature
perturbation, $\mathcal{P_R}(k)$, is given a simple parameterisation
and fitted to the data, together with the background cosmology, using
MCMC likelihood analysis. The spectrum has been described using bins
in wave number $k$
\cite{Bridle:2003sa,Hannestad:2003zs,Bridges:2005br,Bridges:2006zm},
wavelets \cite{Mukherjee:2003ag}, smoothing splines
\cite{Sealfon:2005em} and principal components \cite{Leach:2005av}.
However for MCMC analysis to be feasible, the number of parameters
must be limited, so the reconstructed spectrum has too low a
resolution to reveal anything interesting. By contrast,
`non-parametric' methods {\em assume} the background cosmology
(usually the concordance $\Lambda$CDM model) so that the transfer
function is known, and then invert the data to find the primordial
perturbation spectrum. Methods that have been used are the
Richardson-Lucy deconvolution algorithm
\cite{Shafieloo:2003gf,Shafieloo:2006hs}, an iterative, semi-analytic
process \cite{Kogo:2003yb,Kogo:2004vt}, and a smoothed, least-squares
procedure \cite{Tocchini-Valentini:2004ht,Tocchini-Valentini:2005ja}.
The perturbation spectra produced by the first and third methods have
a prominent step followed by bumps which are rather reminiscent of
decaying oscillations (see Fig.4 in ref.\cite{Hunt:2004vt}). These
features correspond in fact to the depressed quadrupole and the
glitches at low multipole $\ell$.

Encouraged by this we consider possible variations of the primordial
perturbation spectrum beyond the limited set considered so far, and
motivated by the multiple inflation model. We have shown earlier that
a phase transition in a `flat direction' field during inflation (in a
supergravity framework) generates a step in the primordial spectrum
followed by damped oscillations \cite{Hunt:2004vt}. In a physical
supergravity model there are many such flat directions and these will
undergo phase transitions in rapid succession after the first $\sim
10-15$ e-folds of inflation (assuming they all start from the origin
due e.g. to thermal initial conditions) \cite{Adams:1997de}. Hence we
also consider a possible `bump' in the spectrum due to two phase
transitions in rapid succession, which raise and then lower the
inflaton mass. We wish to emphasise that there may well be other
physical frameworks wherein one can expect similar features in the
primordial spectrum. Our intention here is to use a definite and
calculable framework, in order to illustrate how the extraction of
cosmological parameters is dependent on the assumed form of the
primordial spectrum. We find that the precision {\sl WMAP} data can be
fitted just as well {\em without} invoking dark energy if there is
indeed a bump in the primordial spectrum around the position of the
first acoustic peak in the angular power spectrum of the CMB. To fit
the position of the first peak (assuming a flat model as motivated by
inflation) requires however a low Hubble constant, $h \sim 0.44$, in
contrast to the value of $h = 0.72 \pm 0.08$ measured by the Hubble
Key Project ({\sl HKP}) in our local neighbourhood
\cite{Freedman:2000cf}. Although a pure CDM model is well known to
suffer from excess power on small scales, data from the {\sl Sloane
Digital Sky Survey} ({\sl SDSS}) on galaxy clustering
\cite{Tegmark:2003uf} can also be well fitted if there is a $\sim10\%$
component of hot dark matter, as has been noted already using the
2dFGRS data \cite{Elgaroy:2003yh,Blanchard:2003du}. This is
encouraging given the evidence for neutrino mass from oscillations,
and the required value of $\sim0.5$~eV per species is well within the
present experimental upper bound of 2.3~eV \cite{Yao:2006px}. Such a
cold + hot dark matter (CHDM) model with a {\em low} Hubble constant
passes all the usual cosmological tests (e.g. cluster baryon fraction
and $\sigma_8$ from clusters and weak lensing \cite{Blanchard:2003du})
but has difficulty \cite{Blanchard:2005ev} matching the position of
the `baryon acoustic oscillation' (BAO) peak observed in the redshift
two-point correlation function of luminous red galaxies (LRG) in {\sl
SDSS} \cite{Eisenstein:2005su}. We confirm that this is indeed the
case but wish to draw attention to the possibility that this
difficulty may be solved in an {\em inhomogeneous}
Lema\^itre-Tolman-Bondi (LTB) cosmological model wherein we are
located in an underdense void which is expanding faster than the
global rate \cite{Biswas:2006ub}. Such a model may also account for
the SN~Ia Hubble diagram without invoking cosmic acceleration
\cite{Goodwin:1999ej,Celerier:1999hp,Tomita:2000jj,Alnes:2005rw,Enqvist:2006cg,Alnes:2006uk,Celerier:2007jc}.

\section{Multiple Inflation}

In previous work we have discussed the effective potential during
inflation driven by a scalar field in $N=1$ supergravity, which has
couplings to other flat direction fields having gauge and/or Yukawa
couplings \cite{Adams:1997de}. These fields acquire a potential due to
supersymmetry breaking by the large vacuum energy driving inflation
and evolve rapidly to their minima, which are fixed by the
non-renormalisable terms which lift their potential at large field
values. The inflaton's own mass thus jumps as these fields suddenly
acquire large vacuum expectation values (vevs), after having been
trapped at the origin through their coupling to the thermal background
for the first $\sim10-15$ e-folds of inflation. Damped oscillations
are also induced in the inflaton mass as the coupled fields oscillate
in their minima losing energy mainly due to the rapid inflationary
expansion. The resulting curvature perturbation was calculated in our
previous work \cite{Hunt:2004vt} and will be used in this paper as an
input for cosmological parameter extraction using the {\sl WMAP}-3
data.

In order to produce observable effects in the CMB or large-scale
structure the phase transition(s) must take place as cosmologically
relevant scales `exit the horizon' during inflation. There are many
flat directions which can potentially undergo symmetry breaking during
inflation, so it is not unlikely that several phase transitions
occurred in the $\sim 8$ e-folds which is sampled by observations
\cite{Adams:1997de}. The observation that the curvature perturbation
appears to cut off above the scale of the present Hubble radius
suggests that (this last period of) inflation may not have lasted much
longer than the minimum necessary to generate an universe as big as
the present Hubble volume. Whereas this raises a `naturalness' issue,
it is consistent given this state of affairs to consider the
possibility that thermal phase transitions in flat direction fields
$\sim 10-15$ e-folds after the beginning of inflation leave their mark
in the observed scalar density perturbation on the microwave sky and
in the large-scale distribution of galaxies.

\subsection{`Step' model}

The potential for the inflaton $\phi$ and flat direction field $\psi$
(with mass $m$ and $\mu$ respectively) is similar to that given
previously \cite{Hunt:2004vt}:
\begin{equation}
V (\phi, \psi) = \left\{\begin{array}{lc} 
 V_0 - \frac{1}{2}m^2\phi^2, & t < t_1,\\
 V_0 - \frac{1}{2}m^2\phi^2 - \frac{1}{2}\mu^2\psi^2 
     + \frac{1}{2}\lambda\phi^2\psi^2 
     + \frac{\gamma}{M_\mathrm{P}^{n-4}}\psi^{n}, & t \geq t_1. 
                        \end{array}\right.
\label{8ab}
\end{equation}
Here $t_1$ is the time at which the phase transition starts (at $t <
t_1$, $\psi$ is trapped at the origin by thermal effects), $\lambda$
is the coupling between the $\phi$ and $\psi$ fields, and $\gamma$ is
the co-efficient of the leading non-renormalisable operator of order
$n$ which lifts the potential of the flat direction field $\psi$ (and
is determined by the nature of the new physics beyond the effective
field theory description). Note that the quartic coupling above is
generated by a term $\kappa\phi\phi^\dagger\psi^2/M_\mathrm{P}^2$ in
the K\"{a}hler potential with $\kappa \sim 1$ (allowed by symmetry
near $\phi \sim 0$), so $\lambda = \kappa\,H^2/M_\mathrm{P}^2$
\cite{Adams:1997de}. We have not considered non-renormalisable
operators $\propto \phi^{n}/M_\mathrm{P}^{n-4}$ \cite{German:2001tz}
since we are concerned here with the first $\sim10-20$ e-folds of
inflation when $\phi$ is still close to the origin so such operators
are then unimportant for its evolution. We are also not addressing
here the usual `$\eta$-problem' in supergravity models, namely that $m
\sim H$ due to supersymmetry breaking by the vacuum energy driving
inflation. We assume that some mechanism suppresses this mass to
enable sufficient inflation to occur
\cite{Randall:1997kx,Lyth:1998xn}. However this will {\em not} be the
case in general for the flat direction fields, so one would naturally
expect $\mu \sim H$.

Then the change in the inflaton mass-squared after the phase
transition is
\begin{equation} 
\delta m_{\phi}^2 = \lambda\Sigma^2, \quad 
\Sigma = \left[\frac{M_\mathrm{P}^{n-4}}{n\gamma}\left(\mu^2 
 - \lambda\phi^2\right)\right]^{1/\left(n-2\right)} \simeq 
 \left(\frac{\mu^2 M_\mathrm{P}^{n-4}}{n\gamma}\right)^{1/\left(n-2\right)},
\end{equation}
where $\Sigma$ is the vev of the global minimum to which $\psi$
evolves during inflation. The equations of motion are
\begin{eqnarray} 
\ddot{\phi} + 3H\dot{\phi} = & -\frac{\partial V}{\partial\phi} &
= \left(m^2 - \lambda\psi^2\right)\phi, \label{eqm2} \\
\ddot{\psi} + 3H\dot{\psi} = & -\frac{\partial V}{\partial\psi} &
=\left(\mu^2 - \lambda\phi^2 
 - \frac{n\gamma}{M_\mathrm{P}^{n-4}}\psi^{n-2}\right)\psi.
\label{eqm}
\end{eqnarray}

To characterise the comoving curvature perturbation $\mathcal{R}$
\cite{Mukhanov:1990me}, we employ the gauge-invariant quantity
$u=-z\mathcal{R}$, where $z=a\dot{\phi}/H$, $a$ is the cosmological
scale-factor and $H$ is the Hubble parameter during inflation. The
Fourier components of $u$ satisfy the Klein-Gordon equation of motion:
$u''_k + (k^2 - z''/z) u_k = 0$, where the primes indicate derivatives
with respect to conformal time $\eta = \int{\rm d}t/a = -1/aH$ (the
last equality holds in de Sitter space). For convenience, we use the
variable $w_k \equiv \sqrt{2k}u_k$, for which this equation reads:
\begin{equation}
w''_k + w'_k + \left[\widetilde{k}^2\exp\left(-2\widetilde{t}\right) 
 - 2 - \widetilde{m}^2 + \widetilde{\lambda}\widetilde{\psi}^2 
 - \frac{2\widetilde{\lambda}\widetilde{\psi}\widetilde{\psi}'\widetilde{\phi}}{\widetilde{\phi}'}\right] w_k = 0. 
\label{kg}
\end{equation}
where we have used the dimensionless variables: $\widetilde{t}=Ht$,
$\widetilde{\phi}=\phi/M_\mathrm{P}$,
$\widetilde{\psi}=\psi/M_\mathrm{P}$, $\widetilde{m}=m/H$,
$\widetilde{\lambda}=\lambda M_\mathrm{P}^2/H^2$ and
$\widetilde{k}=k/K_0$, where $K_0=a_0 H$ and
$a=a_0\exp(\widetilde{t})$. We also define
$\widetilde{H}=H/M_\mathrm{P}$ and $\widetilde{\mu}=\mu/H$ for
convenience.  Note that now (and henceforth) the dashes refer to
derivatives with respect to $\widetilde{t}$.

We start the integration at an initial value of the scale-factor
$a_0$, which is taken to be several e-folds of inflation before the
phase transition occurs. Similarly we choose an initial value $\phi_0$
for the inflaton field corresponding to several e-folds of inflation
before the phase transition occurs (the precise value does not affect
our results).

We use two parameters to characterise the primordial perturbation
spectrum. The first is the amplitude of the spectrum on large scales
in the slow-roll approximation:
\begin{equation}
\mathcal{P_R}^{(0)} = 
 \left(\frac{H^2}{2\pi\dot{\phi}_0}\right)^2 = 
 \frac{9\widetilde{H}^2}{4\pi^2 \widetilde{m}^4\widetilde{\phi}_0^2}.
\label{PRL}
\end{equation} 
The second is $k_1 \equiv cK_0$ where $c$ is a constant chosen so that
$k_1$ is the position of the step in the spectrum. This can be done
because a mode with wavenumber $k$ `exits the horizon' when the
coefficient of $w_k$ in eq.(\ref{kg}) is zero, hence the wavenumber
$k'_1$ of the mode that exits at the start of the phase transition is
\begin{eqnarray}
k'_1 \equiv & K_0\widetilde{k'}_1 & =
 K_0 \sqrt{2 + \widetilde{m}^2}\ \mathrm{e}^{\widetilde{t}_1},
\end{eqnarray} 
where $\widetilde{t}_1$ is just the number of e-folds of inflation
after which the phase transition occurs. Since we keep
$\widetilde{m}^2$ and $\widetilde{t}_1$ fixed, we have $k'_1 \propto
K_0$. For the phase transition to significantly influence the
primordial perturbation spectrum after the mode with $k = k'_1$ exits the
horizon requires several further e-folds of inflation, depending on
how fast the flat direction evolves. The exponential growth of
$\widetilde{\psi}$ is governed by the value of $\widetilde{\mu}$,
which we also keep fixed. Therefore the position of the step in the
spectrum is proportional to $k_0$. We repeat our analysis for integer
values of $n$, the order of the non-renormalisable term in the
effective field theory description, in the range $12$ to $17$. We set
$\widetilde{\lambda} = \gamma = 1$, $\widetilde{m}^2 = 0.005$,
$\widetilde{\phi}_0 = 0.01$ and $\widetilde{\mu}^2 = 3$. The
fractional change in the inflaton mass-squared due to the phase
transition is
\begin{equation}
\Delta m^2 \equiv \frac{\lambda\Sigma^2}{m^2} = 
 \frac{\widetilde{\lambda}}{\widetilde{m}^2}\left(\frac{\widetilde{\mu}^2
 \widetilde{H}^2}{n\gamma}\right)^{\frac{2}{n-2}}.
\end{equation}
Thus fixing the value of $n$ does not entirely fix $\Delta m^2$,
because from eq.(\ref{PRL}) varying $P_{\mathcal{R}}^{(0)}$ also alters
$\widetilde{H}$. A typical value is $P_{\mathcal{R}}^{(0)} \sim
10^{-9}$ so the corresponding Hubble parameter is $H \sim 3 \times
10^{-8} M_\mathrm{P}$ i.e. an inflationary energy scale of $\sim 2 \times
10^{14}$ GeV. This is comfortably within the upper limit of $\sim 2
\times 10^{16}$ GeV set by the {\sl WMAP} bound on inflationary
gravitational waves \cite{Kinney:2006qm}. Indeed the gravitational
wave background is completely negligible for the `new' inflation
potential we consider, with the tensor to scalar ratio expected to be:
$r \sim \widetilde{m}^4\widetilde{\phi}^2 \sim 10^{-9}$.

We varied four parameters describing the homogeneous background
cosmology which is taken to be a spatially flat
Friedman-Robertson-Walker (FRW) universe: the physical baryon density
$\omega_\mathrm{b} \equiv \Omega_\mathrm{b}h^2$, the physical cold
dark matter density $\omega_\mathrm{c} \equiv \Omega_\mathrm{c}h^2$,
the ratio $\theta$ of the sound horizon to the angular diameter
distance (multiplied by 100), and the optical depth $\tau$ (due to
reionisation) to the last scattering surface. The dark energy density
is given by $\Omega_\Lambda = 1 - \Omega_\mathrm{m}$, where
$\Omega_\mathrm{m} \equiv \Omega_{\rm c} + \Omega_{\rm b}$.

\subsection{`Bump' model}

Next we consider a multiple inflation model with two successive phase
transitions caused by 2 flat direction fields, $\psi_1$ and $\psi_2$.
The potential is now:
\begin{equation}
V \left(\phi, \psi_1, \psi_2\right) =\left\{\begin{array}{ll}
V_0 - \frac{1}{2}m^2 \phi^2, & t < t_1,\\
V_0 - \frac{1}{2}m^2\phi^2 - \frac{1}{2}\mu_1^2\psi_1^2 
 + \frac{1}{2}\lambda_1\phi^2\psi_1^2 
 + \frac{\gamma_1}{M_\mathrm{P}^{n_1 - 4}}\psi^{n_1}, 
& t_2 \geq t \geq t_1,\\
V_0 - \frac{1}{2}m^2\phi^2 - \frac{1}{2}\mu_1^2\psi_1^2 
+ \frac{1}{2}\lambda_1\phi^2\psi_1^2
+\frac{\gamma_1}{M_{P}^{n_1 -4}}\psi^{n_1}\\
- \frac{1}{2}\mu_2^2\psi_2^2 - \frac{1}{2}\lambda_2\phi^2\psi_2^2
+ \frac{\gamma_2}{M_\mathrm{P}^{n_2 - 4}}\psi^{n_2}, & t \geq t_2,
\end{array}\right.
\end{equation}
where $t_1$ and $t_2$ are the times at which the first and second
phase transitions occur. After $t_2$, for example, the equations of
motion are
\begin{eqnarray}
\ddot{\phi} + 3H\dot{\phi} = & -\frac{\partial V}{\partial\phi} & 
 = \left(m^2 - \lambda_1\psi_1^2 + \lambda_2\psi_2^2\right)\phi,\\
\ddot{\psi_1} + 3H\dot{\psi_1} = & -\frac{\partial V}{\partial\psi_1} & 
 = \left(\mu_1^2 - \lambda_1\phi^{2} 
 - \frac{n_1\gamma_1}{M_\mathrm{P}^{n_1 - 4}}\psi_1^{n_1 - 2}\right)\psi_1,\\
\ddot{\psi_2} + 3H\dot{\psi_2} = & -\frac{\partial V}{\partial\psi_2} & 
 = \left(\mu_2^2 + \lambda_2\phi^2 
 - \frac{n_2\gamma_2}{M_\mathrm{P}^{n_2 - 4}}\psi_2^{n_2 - 2}\right)\psi_2,
\end{eqnarray}
and
\begin{equation}
\frac{1}{z}\frac{d^{2}z}{d\eta^{2}} = a^{2}\left(2H^{2} + m^{2}
 - \lambda_1\psi_1^{2} + \lambda_2\psi_2^{2} 
 - \frac{2\lambda_1\psi_1\dot{\psi_1}\phi}{\dot{\phi}}
 + \frac{2\lambda_2\psi_2\dot{\psi_2}\phi}{\dot{\phi}}\right).
\end{equation}
The inflaton mass $m_\phi$ thus changes due to the phase transitions as
\begin{equation}
m^2 \rightarrow m^2 - \lambda_1\Sigma_1^2\rightarrow m^2 
 - \lambda_1\Sigma_1^2 + \lambda_{2}\Sigma_2^2.
\end{equation}
By choosing $\lambda_2$ to be of opposite sign to $\lambda_1$
(possible in the supergravity model), a bump is thus generated in
$\mathcal{P_R}$. In the slow-roll approximation the amplitude of the
primordial perturbation spectrum first increases from
$\mathcal{P_R}^{(0)}$ to $\mathcal{P_R}^{(1)}$ then falls to
$\mathcal{P_R}^{(2)}$, moving from low to high wavenumbers, where
\begin{equation}
\mathcal{P_R}^{(0)} = 
\frac{9\widetilde{H}^{2}}{4\pi^{2}\widetilde{m}^{4}
\widetilde{\phi}_{0}^{2}}, \qquad 
\mathcal{P_R}^{(1)} = 
\frac{\mathcal{P}_{\mathcal{R}}^{\left(0\right)}}{\left(1-\Delta
  m_{1}^{2}\right)^{2}}, \qquad 
\mathcal{P_R}^{(2)} = 
 \frac{\mathcal{P_R}^{(0)}}{\left(1 - \Delta m_{1}^{2} + 
 \Delta m_{2}^2\right)^2}\ .
\label{ampl}
\end{equation} 
We calculate the curvature perturbation spectrum for this model in a
way similar to that for the model with one phase transition. As before
we specify the homogeneous background cosmology using $\omega_{\rm
b}$, $\theta$ and $\tau$ and, anticipating the discussion later, the
fraction of dark matter in the form of neutrinos $f_\nu \equiv
\Omega_{\nu}/\Omega_{\rm d}$ where the total dark matter density is
$\Omega_{\rm d} \equiv \Omega_{\rm c} + \Omega_{\nu}$. The primordial
perturbation spectrum is parameterised in a similar way to that of the
first model using $\mathcal{P_R}^{(0)}$ and $k_1 \equiv cK_0$, which
is now the approximate position of the first step in the spectrum. A
measure of the position of the second step is given by the third
parameter
\begin{equation}
k_2 \equiv k_1 \mathrm{e}^{(\widetilde{t}_2 - \widetilde{t}_1)},
\label{kpost}
\end{equation}
where the exponent is just the number of Hubble times after the
beginning of the first phase transition when the second one starts.
Initially we examined models with $n_1=12$ and $n_2=13$, and then we
let $\mathcal{P_R}^{(1)}$ and $\mathcal{P_R}^{(2)}$ vary freely.  We
set $\widetilde{\lambda}_1$, $\widetilde{\lambda}_2$, $\gamma_1$ and
$\gamma_2$ all equal to unity, $\widetilde{m}^2=0.005$,
$\widetilde{\phi}_0=0.01$ and $\widetilde{\mu}_1^2 =
\widetilde{\mu}_2^2 = 3$ throughout.

\section{The Data Sets}

We fit to the {\sl WMAP} 3-year \cite{Hinshaw:2006ia}
temperature-temperature (TT), temperature-electric polarisation (TE),
and electric-electric polarisation (EE) spectra alone as we wish to
avoid possible systematic problems associated with combining other CMB
data sets. We also fit the linear matter power spectrum
$\mathcal{P}_\mathrm{m} (k)$ to the {\sl SDSS} measurement of the real
space galaxy power spectrum $\mathcal{P}_\mathrm{g}(k)$
\cite{Tegmark:2003uf}. The two spectra are taken to be related through
a {\em scale-independent} bias factor $b_{\rm SDSS}$ so that
$\mathcal{P}_\mathrm{m}(k) = b_{\rm SDSS}^2
\mathcal{P}_\mathrm{g}(k)$. This bias is expected to be close to unity
and we analytically marginalise over it using a flat prior.

We also fit our models to the redshift two-point correlation function
of the {\sl SDSS} LRG sample, which was obtained assuming a fiducial
cosmological model with $\Omega_\mathrm{m} = 0.3$, $\Omega_{\Lambda} =
0.7$ and $h=0.7$ \cite{Eisenstein:2005su}. The correlation function is
dependent on the choice of background cosmology so we rescale it
appropriately as explained below, in order to confront it with the
other cosmological models we consider. Again we consider a linear bias
$b_\mathrm{LRG}$, and determine it from the fit itself.

\begin{table}
\begin{center}
\begin{tabular}{||l|l|l||}
\hline
\hline
Parameter & Lower limit & Upper limit  \\
\hline
\hline
$\omega_\mathrm{b}$ & $0.005$ & $0.1$  \\
\hline
$\omega_\mathrm{c}$ & $0.01$ & $0.99$  \\
\hline
$\theta$ & $0.5$ & $10.0$  \\
\hline
$\tau$ & $0.01$ & $0.8$  \\
\hline
$\ln\left(10^{10}\mathcal{P_R}^{(0)}\right)$ & $0.01$ & $6.0$ \\
\hline
$10^4 k_1/\mathrm{Mpc}^{-1}$ & $0.01$ & $600.0$  \\
\hline
$b_\mathrm{LRG}$ & $1.0$ & $4.0$  \\
\hline
$h$ & $0.4$ & $1.0$  \\
\hline
Age/Gyr & $10.0$ & $20.0$  \\
\hline
\hline
\end{tabular}
\end{center}
\caption{\label{tab:priors} 
  The priors adopted on the input parameters
  of the step model, as well as on the derived parameters: the
  Hubble constant and the age of the Universe.}
\end{table}

\begin{table}
\begin{center}
\begin{tabular}{||l|l|l||}
\hline
\hline
Parameter & Lower limit & Upper limit  \\
\hline
\hline
$\omega_\mathrm{b}$ & $0.005$ & $0.1$  \\
\hline
$\theta$ & $0.5$ & $10.0$  \\
\hline
$\tau$ & $0.01$ & $0.8$  \\
\hline
$f_\nu$ & $0.01$ & $0.3$  \\
\hline
$\ln\left(10^{10}\mathcal{P_R}^{(0)}\right)$ & $0.01$ & $6.0$ \\
\hline
$b_\mathrm{LRG}$ & $1.0$ & $4.0$  \\
\hline
$h$ & $0.1$ & $1.0$  \\
\hline
Age/Gyr & $10.0$ & $20.0$  \\
\hline
\hline
\end{tabular}
\end{center}
\caption{\label{tab:priors2} 
  The priors adopted on the input parameters of the bump model with
  $n_1 = 12$, $n_1 = 13$, as well as on the derived parameters: the
  Hubble constant and the age of the Universe. The priors set on the
  position of the steps are given in Table~\ref{tab:priors3}.}
\end{table} 

\begin{table}
\begin{center}
\begin{tabular}{|c|c|c|c|c|}
\hline 
&
WMAP&
+{\sl SDSS}&
+LRG&
+{\sl SDSS}+LRG\\
\hline
\hline 
$10^4 k_1/\mathrm{Mpc}^{-1}$&
$600.0$&
$600.0$&
$600.0$&
$600.0$\\
\hline 
$10^4 k_2/\mathrm{Mpc}^{-1}$&
$1100.0$&
$1100.0$&
$600.0$&
$600.0$\\
\hline
\end{tabular}
\end{center}
\caption{\label{tab:priors3} 
The prior upper limits set for the position of the steps in the
 spectrum for the `bump' model. In each case the lower limit is
 $10^{-6}\,\mathrm{Mpc}^{-1}$.}
\end{table}

\subsection{\sl WMAP}

To date the most accurate observations of the CMB angular power
spectra, both TT and TE, have been made by the {\sl WMAP} satellite
using 20 differential radiometers arranged in 10 `differencing
assemblies' at 5 different frequency bands between 23 and 94 GHz. Each
differencing assembly produces a statistically independent stream of
time-ordered data. Full sky maps were generated from the calibrated
data using iterative algorithms. The use of different frequency
channels enable astrophysical foreground signals to be removed and the
angular power spectra were calculated using both `pseudo-$C_{\ell}$'
and maximum likelihood based methods \cite{Hinshaw:2006ia}. We use the
3-year data release of the TT, TE and EE spectra and also the code for
calculating the likelihood (incorporating the covariance matrix), made
publicly available by the {\sl WMAP}
team.\footnote{http://lambda.gsfc.nasa.gov/}

Our special interest is in the fact that apart from the quadrupole
there are several other multipoles for which the binned power lies
outside the 1$\sigma$ cosmic variance error. These are associated with
small, sharp features in the TT spectrum termed glitches, e.g. at
$\ell$ = 22, 40 and 120 in both the 1-year and 3-year data release.
The pseudo-$C_{\ell}$ method produces correlated estimates for
neighbouring $C_{\ell}$'s which means it is difficult to judge the
goodness-of-fit of a model by eye.  The contribution to the $\chi^2$
per multipole for the best fit $\Lambda$CDM model is shown in Fig.17
of ref.\cite{Hinshaw:2006ia}, where it is apparent that the excess
$\chi^2$ originates largely from the multipoles at $\ell\lesssim120$.

\subsection{\sl SDSS}

The Sloan Digital Sky Survey ({\sl SDSS}) consists of a 5 band imaging
survey and a spectroscopic survey covering a large fraction of the
sky, carried out together using a 2.5 m wide-field telescope. For the
spectroscopic survey targets are chosen from the imaging survey to
produce three catalogues: the main galaxy sample, the luminous, red
galaxy (LRG) sample and the quasar sample, which extend out to
redshifts of $z \sim 0.3$, 0.5 and 5 respectively. The matter power
spectrum (up to a multiplicative bias factor $b_\mathrm{SDSS}$) has
been measured using about $2 \times 10^5$ galaxies which form the
majority of the main galaxy sample from the second {\sl SDSS} Data
Release \cite{Tegmark:2003uf}. After correcting for redshift-space
distortion due to galaxy peculiar velocities, the galaxy density field
was expanded in terms of Karhunen-Loeve eigenmodes (which is more
convenient than a Fourier expansion for surveys with complex
geometries) and the power spectrum was estimated using a quadratic
estimator. We fit to the 19 k-band measurements in the range 
$0.01 < k < 0.2\,h\,\mathrm{Mpc^{-1}}$, as the fluctuations
turn non-linear on smaller scales.

\subsection{LRG}

The large effective volume of the {\sl SDSS} LRG sample allowed an
unambiguous detection of the BAO peak in the two-point correlation
function of galaxy clustering \cite{Eisenstein:2005su}. The redshifts
of $5 \times 10^4$ LRG's were translated into comoving coordinates
assuming a fiducial $\Omega_\Lambda = 0.7, \Omega_\mathrm{m} = 0.3$
cosmology. The correlation function was then found using the
Land-Szalay estimator. This can be rescaled for other
cosmological models as follows.

An object at redshift $z$ with an angular size of $\Delta\theta$ and
extending over a redshift interval $\Delta z$ has dimensions in
redshift space $x_{\|}$ and $x_{\bot}$, parallel and perpendicular to
the line-of-sight, given by
\begin{equation}
x_{\|} = \frac{\Delta z}{H\left(z\right)}, \quad
x_{\bot} = \left(1+z\right)D_{A}\left(z\right)\Delta\theta.
\end{equation}
Thus the volume in redshift space is $\propto D_\mathrm{A}^2(z)/H(z)$,
where $D_\mathrm{A}$ is the angular diameter distance. The distances
between galaxies in redshift space, hence the scale of features in the
redshift correlation function, are proportional to the cube root of
the volume occupied by the galaxies in redshift space. Therefore the
multiplicative rescaling factor between the scale of the BAO peak in
the fiducial model and that in another model is just:
\begin{equation}
\gamma_{\mathrm{res}} \equiv
\left[\frac{D_\mathrm{A}^2(z) H_{\mathrm{fid}}(z)}{D_{\mathrm{A},
 \mathrm{fid}}^2(z) H(z)}\right]^{\frac{1}{3}},
\label{scale}
\end{equation}
i.e.  $x_{\mathrm{peak}} = \gamma_{\mathrm{res}} \times
x_{\mathrm{peak},\mathrm{fid}}$. After rescaling in this way for the
cosmological models we consider, the $\chi^2$ statistic for the fit of
$\xi$ to the data points was found using the appropriate covariance
matrix.\footnote{http://cmb.as.arizona.edu/~eisenste/}

\section{Calculation of the observables}

Since MCMC analysis of inflation requires the evaluation of the
observables for typically tens of thousands of different choices of
cosmological parameters, it is important that each calculation be as
fast as possible.

The inflaton mass is constant before and after the phase
transition(s), so the equations of motion for $\widetilde{\phi}$ and
$w_{k}$ have the solutions
\begin{eqnarray}
\widetilde{\phi} & = & c_1\mathrm{e}^{\widetilde{t}r_{+}} 
 + c_2\mathrm{e}^{\widetilde{t}r_{-}}, \label{9ab}\\
w_{k} & = & \mathrm{e}^{-\widetilde{t}/2} \left[
   c_3\mathrm{H}_\nu^{(1)} (\widetilde{k}\mathrm{e}^{-\widetilde{t}}) 
 + c_4\mathrm{H}_\nu^{(2)}(\widetilde{k}\mathrm{e}^{-\widetilde{t}})\right].
\label{9aa}
\end{eqnarray}
Here $c_{1}$ to $c_{4}$ are constants,
$\mathrm{H}_\nu^{\left(1\right)}$ and
$\mathrm{H}_\nu^{\left(2\right)}$ are Hankel functions of order $\nu =
\sqrt{\frac{9}{4} - \frac{m_\phi^2}{H^2}}$, and $r_{\pm} = -3/2 \pm
\nu.$ We evolve $\widetilde{\phi}$, $\widetilde{\psi}$ and $w_k$
numerically through the phase transition(s), matching to
eqs.(\ref{9ab}) and (\ref{9aa}). The slow-roll approximation is
applicable well before the phase transition(s), so the initial values
are related as
$\widetilde{\phi}_{0}'=(\widetilde{m}^2/3)\widetilde{\phi}_{0}$.  We
begin the integration of the Klein-Gordon equation at
$\widetilde{t}_\mathrm{start}$ when the condition
\begin{equation}
\epsilon k^2 = \frac{1}{z} \frac{\mathrm{d}^2 z}{\mathrm{d}\eta^2}
\end{equation}
is satisfied. The arbitrary constant $\epsilon$ is chosen to be
$5\times 10^{-5}$ which is sufficiently small for $\mathcal{P_R}$ to
be independent of the precise value of $\widetilde{t}_\mathrm{start}$,
yet large enough for the numerical integration to be feasible. For the
Bunch-Davies vacuum, the initial conditions for $w_k$ are
\cite{Mukhanov:1990me}
\begin{equation}
 w_{k} (\widetilde{t}_\mathrm{start}) = 1, \quad
 w'_{k}(\widetilde{t}_\mathrm{start}) = 
  -i \widetilde{k}\exp\left(-\widetilde{t}_\mathrm{start}\right).
\end{equation}

The amplitude of the primordial perturbation spectrum on large scales
was given in eq.(\ref{PRL}); on smaller scales it is
\begin{equation}
\mathcal{P_R} = \frac{k^{3}}{2\pi^{2}}\frac{\left|u_k\right|^2}{z^2} =
\frac{\widetilde{k}^2\widetilde{H}^2}{4\pi^2\widetilde{\phi}^{\prime 2}
\exp\left(2\widetilde{t}\right)}\left|w_k\right|^2,
\end{equation}
evaluated when the mode has crossed well outside the horizon. Rather
than calculating $\mathcal{P}_{\mathcal{R}}$ in this way for every
wavenumber, we use cubic spline interpolation between
$\mathcal{O}\left(100\right)$ wavenumbers $k_i$ to save time. These
sample the spectrum more densely where its curvature is higher, in
order to minimise the interpolation error. The $k_i$ values are found
each time using the adaptive sampling algorithm described in
Appendix~\ref{sampling}.

Taking the interpolated primordial perturbation spectrum as input, we
use a modified version of the cosmological Boltzmann code CAMB
\cite{Lewis:1999bs} to evaluate the CMB TT, TE and EE spectra, as well
as the matter power spectrum.\footnote{http://camb.info}

We find the two-point correlation function using the following
procedure: for $10^{-5}\leq k \leq 2\,h\,\mathrm{Mpc^{-1}}$ the linear
matter power spectrum $P_\mathrm{m}(k)$ is obtained using CAMB, while
outside this range baryon oscillations are negligible, and so to
calculate the matter power spectrum $P_\mathrm{m}(k)$ for $10^{-6}
\leq k \leq 10^{-5}\,h\,\mathrm{Mpc^{-1}}$ and $2 \leq k \leq
10^{2}\,h\,\mathrm{Mpc^{-1}}$ we use the no-baryon transfer function
fitting formula of ref.\cite{Eisenstein:1997ik} normalised to the CAMB
transfer function. This significantly increases the speed of the
computation. The `Halofit' procedure \cite{Smith:2002dz} is then used
to apply the corrections from non-linear evolution to produce the
non-linear matter power spectrum
$P_\mathrm{m}^\mathrm{NL}\left(k\right)$. The LRG power spectrum is
given by $P_\mathrm{LRG}(k) = b_\mathrm{LRG}^2
P_\mathrm{m}^\mathrm{NL}(k)$. The real space correlation function
$\xi_{\rm c}(x)$ is then calculated by taking the Fourier transform.
The integral is performed over a finite range of $k$ using the FFTLog
code \cite{Hamilton:1999uv} which takes the fast Fourier transform of
a series of logarithmically spaced points. The wavenumber range must
be broad enough to prevent ringing and aliasing effects from the FFT
over the interval of $x$ for which we require $\xi_{\rm c}$. We find
the range $10^{-6} \leq k \leq 10^{2}\,h\,\mathrm{Mpc^{-1}}$ to be
sufficient. Finally, the angle averaged redshift correlation function
is found from the real space correlation function using
\begin{equation}
\xi(x)  =\left(1 + \frac{2}{3}f + \frac{1}{5}f^2\right)
\xi_{\rm c}(x),\qquad f \equiv \frac{\Omega_\mathrm{m}^{3/5}}{b_{\rm LRG}},
\end{equation}
which corrects for redshift space distortion effects
\cite{Kaiser:1987qv,Hamilton:1991es}.

\section{Results}
 
We calculate the mean values of the marginalised cosmological
parameters together with their $68$\% confidence limits, for the four
combinations of data sets: {\sl WMAP}, {\sl WMAP + SDSS}, {\sl WMAP} +
LRG, {\sl WMAP + SDSS} + LRG. In Table~\ref{tab:flat} we show this for
the $\Lambda$CDM model with a Harrison-Zeldovich spectrum. Adding an
overall `tilt' to the spectrum creates a noticeable improvement (by
$\Delta\chi^2 \sim 7-9$) as is seen from Table~\ref{tab:tilt} which
gives results for the $\Lambda$CDM model with a power-law scalar
spectral index of $n_\mathrm{s} \simeq 0.95$ (pivot point $k =
0.05\,\mathrm{Mpc}^{-1}$). The $\chi^2$ goodness-of-fit statistic is
calculated using the {\sl WMAP} likelihood
code, supplemented by our
estimate for the LRG fits obtained using the covariance matrix used in
ref.\cite{Eisenstein:2005su}.

In order to quantify the desirable compromise between improving the
fit and adding new parameters, it has become customary to use the
``Akaike information criterion'' defined as $\mathrm{AIC} \equiv
-2\ln{\cal L}_\mathrm{max} + 2 N$ \cite{Akaike:1974}, where ${\cal
L}_\mathrm{max}$ is the maximum likelihood and $N$ the number of
parameters. Although not a substitute for a full Bayesian evidence
calculation (which is beyond the scope of this work), this is a
commonly used guide for judging whether additional parameters are
justified --- the model with the {\em minimum} AIC value is in some
sense preferred \cite{Liddle:2004nh}. According to this criterion, the
power-law $\Lambda$CDM model is preferred over the scale-invariant
$\Lambda$CDM model, hence we consider the former to be the benchmark
against which our models should be judged. However we wish to
emphasise that we have {\em physical} motivation for the non-standard
primordial spectra we consider. We are not performing a pure parameter
fitting exercise for which criteria like the AIC might be more
relevant.

\begin{table}
\begin{center}
\begin{tabular}{|c|c|c|c|c|}
\hline 
&
{\sl WMAP}&
+{\sl SDSS}&
+LRG&
+{\sl SDSS}+LRG\\
\hline
\hline 
$\Omega_\mathrm{b}h^2$&
$0.02381_{-0.00043}^{+0.00042}$&
$0.02393_{-0.00039}^{+0.00039}$&
$0.02395_{-0.00038}^{+0.00037}$&
$0.02394_{-0.00038}^{+0.00038}$\\
\hline 
$\Omega_\mathrm{c}h^2$&
$0.1031_{-0.0079}^{+0.0077}$&
$0.1131_{-0.0052}^{+0.0054}$&
$0.1163_{-0.0045}^{+0.0046}$&
$0.1173_{-0.0041}^{+0.0041}$\\
\hline 
$\theta$&
$1.0451_{-0.0027}^{+0.0029}$&
$1.0463_{-0.0028}^{+0.0027}$&
$1.0465_{-0.0028}^{+0.0029}$&
$1.0465_{-0.0026}^{+0.0027}$\\
\hline 
$\tau$&
$0.139_{-0.013}^{+0.013}$&
$0.129_{-0.012}^{+0.014}$&
$0.123_{-0.011}^{+0.012}$&
$0.120_{-0.012}^{+0.012}$\\
\hline
$\ln\left(10^{10}\mathcal{P_{R}}\right)$&
$3.137_{-0.052}^{+0.055}$&
$3.160_{-0.049}^{+0.050}$&
$3.161_{-0.049}^{+0.051}$&
$3.158_{-0.047}^{+0.053}$\\
\hline
$b_\mathrm{LRG}$&
&
&
$2.123_{-0.091}^{+0.089}$&
$2.129_{-0.093}^{+0.093}$\\
\hline
\hline
$\Omega_{\Lambda}$&
$0.785_{-0.029}^{+0.029}$&
$0.746_{-0.022}^{+0.021}$&
$0.732_{-0.018}^{+0.018}$&
$0.728_{-0.017}^{+0.017}$\\
\hline
Age/Gyr&
$13.403_{-0.090}^{+0.091}$&
$13.423_{-0.091}^{+0.091}$&
$13.432_{-0.094}^{+0.092}$&
$13.438_{-0.084}^{+0.088}$\\
\hline
$\Omega_\mathrm{b}$&
$0.215_{-0.029}^{+0.029}$&
$0.254_{-0.021}^{+0.022}$&
$0.268_{-0.018}^{+0.018}$&
$0.272_{-0.017}^{+0.017}$\\
\hline
$\sigma_8$&
$0.801_{-0.050}^{+0.049}$&
$0.856_{-0.032}^{+0.033}$&
$0.870_{-0.031}^{+0.030}$&
$0.873_{-0.029}^{+0.029}$\\
\hline
$z_\mathrm{reion}$&
$14.5_{-2.0}^{+2.0}$&
$14.1_{-2.1}^{+2.0}$&
$13.7_{-1.9}^{+2.1}$&
$13.5_{-2.0}^{+2.2}$\\
\hline
$h$&
$0.772_{-0.029}^{+0.030}$&
$0.736_{-0.020}^{+0.018}$&
$0.724_{-0.015}^{+0.014}$&
$0.721_{-0.014}^{+0.014}$\\
\hline
\hline
$\chi^2$&
$11260$&
$11279$&
$11284$&
$11301$\\
\hline
$\Delta_\mathrm{AIC}$&
$5$&
$6$&
$7$&
$7$\\
\hline
\end{tabular}
\end{center}
\caption{\label{tab:flat} $1 \sigma$ constraints on the marginalised
 cosmological parameters for a scale-invariant $\Lambda$CDM model. The
 6 parameters in the upper part of the Table are varied by CosmoMC,
 while those in the lower part are derived quantities. The $\chi^2$ of
 the fit is given, as is the Akaike information criterion
 relative to the power-law $\Lambda$CDM model in
 Table~\ref{tab:tilt}.}
\end{table}

\begin{table}
\begin{center}
\begin{tabular}{|c|c|c|c|c|}
\hline 
&
{\sl WMAP}&
+{\sl SDSS}&
+LRG&
+{\sl SDSS}+LRG\\
\hline
\hline 
$\Omega_\mathrm{b} h^2$&
$0.02219_{-0.00070}^{+0.00070}$&
$0.02236_{-0.00069}^{+0.00066}$&
$0.02229_{-0.00068}^{+0.00069}$&
$0.02240_{-0.00067}^{+0.00064}$\\
\hline 
$\Omega_\mathrm{c} h^2$&
$0.1052_{-0.0084}^{+0.0084}$&
$0.1147_{-0.0054}^{+0.0056}$&
$0.1147_{-0.0044}^{+0.0043}$&
$0.1168_{-0.0038}^{+0.0038}$\\
\hline 
$\theta$&
$1.0391_{-0.0037}^{+0.0038}$&
$1.0406_{-0.0036}^{+0.0036}$&
$1.0403_{-0.0035}^{+0.0035}$&
$1.0410_{-0.0033}^{+0.0033}$\\
\hline 
$\tau$&
$0.089_{-0.014}^{+0.014}$&
$0.083_{-0.013}^{+0.014}$&
$0.080_{-0.013}^{+0.013}$&
$0.080_{-0.013}^{+0.013}$\\
\hline
$n_\mathrm{s}$&
$0.9537_{-0.0081}^{+0.0072}$&
$0.9542_{-0.0070}^{+0.0075}$&
$0.9527_{-0.0075}^{+0.0072}$&
$0.9547_{-0.0072}^{+0.0068}$\\
\hline
$\ln\left(10^{10}\mathcal{P_R}\right)$&
$3.017_{-0.065}^{+0.065}$&
$3.046_{-0.062}^{+0.061}$&
$3.039_{-0.062}^{+0.062}$&
$3.048_{-0.060}^{+0.060}$\\
\hline
$b_\mathrm{LRG}$&
&
&
$2.26_{-0.10}^{+0.10}$&
$2.25_{-0.10}^{+0.10}$\\
\hline
\hline
$\Omega_\Lambda$&
$0.758_{-0.037}^{+0.037}$&
$0.716_{-0.027}^{+0.028}$&
$0.716_{-0.020}^{+0.020}$&
$0.708_{-0.018}^{+0.019}$\\
\hline
Age/Gyr&
$13.75_{-0.17}^{+0.16}$&
$13.75_{-0.15}^{+0.15}$&
$13.77_{-0.15}^{+0.15}$&
$13.75_{-0.14}^{+0.14}$\\
\hline
$\Omega_\mathrm{m}$&
$0.242_{-0.037}^{+0.037}$&
$0.284_{-0.028}^{+0.027}$&
$0.284_{-0.020}^{+0.020}$&
$0.292_{-0.019}^{+0.018}$\\
\hline
$\sigma_8$&
$0.754_{-0.050}^{+0.049}$&
$0.804_{-0.034}^{+0.035}$&
$0.801_{-0.035}^{+0.036}$&
$0.813_{-0.032}^{+0.031}$\\
\hline
$z_\mathrm{reion}$&
$11.1_{-2.4}^{+2.5}$&
$10.8_{-2.4}^{+2.5}$&
$10.5_{-2.5}^{+2.5}$&
$10.5_{-2.6}^{+2.6}$\\
\hline
$h$&
$0.730_{-0.034}^{+0.034}$&
$0.697_{-0.023}^{+0.024}$&
$0.695_{-0.017}^{+0.017}$&
$0.691_{-0.016}^{+0.016}$\\
\hline
\hline
$\chi^2$&
$11253$&
$11271$&
$11275$&
$11292$\\
\hline
$\Delta_\mathrm{AIC}$&
$0$&
$0$&
$0$&
$0$\\
\hline
\end{tabular}
\end{center}
\caption{\label{tab:tilt}
 $1\sigma$ constraints on the marginalised cosmological parameters for
 the power-law $\Lambda$CDM model. The 6 parameters in the upper part
 of the Table are varied by CosmoMC, while those in the lower part are
 derived quantities. The $\chi^2$ of the fit is given as is the
 relative Akaike information criterion (normalised to be zero for this
 model).}
\end{table}

\subsection{$\Lambda$CDM `step' model}

In Tables~\ref{tab:wmap1step} to \ref{tab:all1step} we show that
adding a step to the spectrum improves the fit by $\Delta\chi^2 \sim
1-2$ relative to the scale-invariant model, but not as much as addition
of a tilt. Since the model has 2 additional parameters --- the position
of the step $k_1$ and $n$ (which determines $\Delta m^2$) ---
$\Delta_\mathrm{AIC}$ is {\em positive} relative to the power-law
$\Lambda$CDM model.

\begin{figure}[tbh]
\includegraphics[angle=0,scale=0.85]{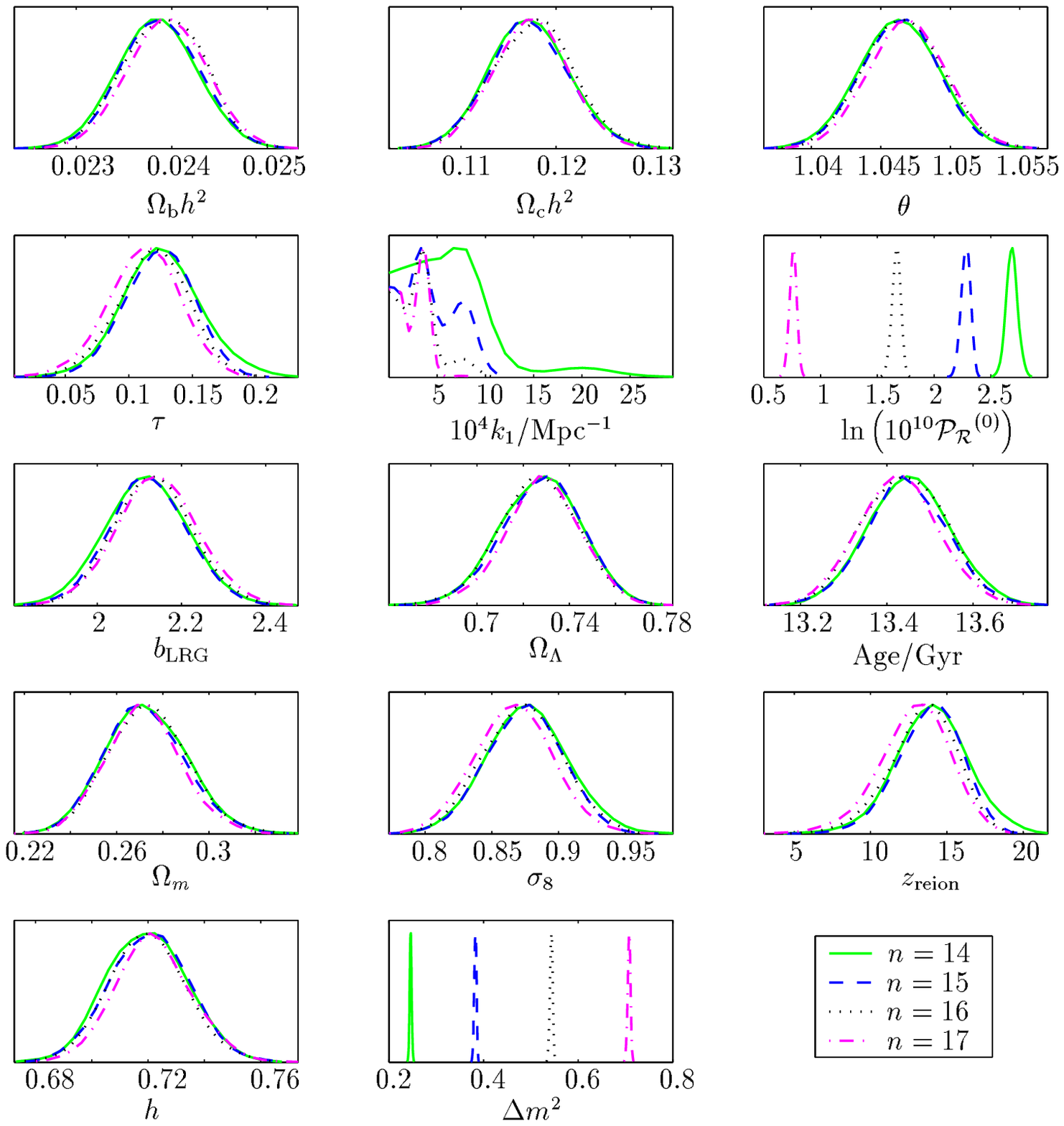}
\caption{\label{all1Lmany2a}
 1-dimensional likelihood distributions of the marginalised
 cosmological parameters for the $\Lambda$CDM step model using the
 {\sl WMAP} + {\sl SDSS} + LRG data.}
\end{figure}

We have not shown the results for the $\Lambda$CDM step model allowing
an overall tilt in the spectrum; although the fits do improve again
(by $\Delta\chi^2 \sim1-2$) relative to the power-law $\Lambda$CDM
model, they are not favoured by the AIC because of the 2 additional
parameters $k_1$ and $n$. The values of the background cosmological
parameters do not change significantly with $n$ either, and remain
similar to those of the power-law $\Lambda$CDM model.  However, the
values of both $\mathcal{P_{R}}^{\left(1\right)}$ and $k_1$ fall with
increasing $n$, as seen in the 1-D likelihood distributions for the
parameters shown in Fig.\ref{all1Lmany2a}.

\begin{table}
\begin{center}
\begin{tabular}{|c|c|c|c|c|c|c|}
\hline 
$n$&
12&
13&
14&
15&
16&
17\\
\hline
\hline 
$\Omega_\mathrm{b} h^2$&
$0.02366_{-0.00043}^{+0.00042}$&
$0.02366_{-0.00046}^{+0.00045}$&
$0.02363_{-0.00046}^{+0.00046}$&
$0.02362_{+0.00045}^{+0.00045}$&
$0.02370_{-0.00044}^{+0.00044}$&
$0.02371_{-0.00045}^{+0.00046}$\\
\hline 
$\Omega_\mathrm{c} h^2$&
$0.1030_{-0.0079}^{+0.0083}$&
$0.1022_{-0.0082}^{+0.0084}$&
$0.1002_{-0.0086}^{+0.0083}$&
$0.1002_{-0.0083}^{+0.0080}$&
$0.1006_{-0.0084}^{+0.0085}$&
$0.1017_{-0.0087}^{+0.0086}$\\
\hline 
$\theta$&
$1.0446_{-0.0029}^{+0.0030}$&
$1.0445_{-0.0031}^{+0.0030}$&
$1.0444_{-0.0030}^{+0.0029}$&
$1.0445_{-0.0029}^{+0.0030}$&
$1.0450_{-0.0031}^{+0.0031}$&
$1.0451_{-0.0030}^{+0.0029}$\\
\hline 
$\tau$&
$0.151_{-0.014}^{+0.014}$&
$0.157_{-0.016}^{+0.016}$&
$0.159_{-0.017}^{+0.015}$&
$0.148_{-0.015}^{+0.013}$&
$0.142_{-0.012}^{+0.013}$&
$0.129_{-0.012}^{+0.014}$\\
\hline
$10^4 k_1/\mathrm{Mpc}^{-1}$&
$20.8_{-7.1}^{+4.8}$&
$14.9_{-6.2}^{+5.6}$&
$9.5_{-3.4}^{+0.7}$&
$6.1_{-2.3}^{+1.8}$&
$4.9_{-1.9}^{+2.5}$&
$3.2_{-1.6}^{+0.4}$\\
\hline
$\ln\left(10^{10}\mathcal{P_R}^{(0)}\right)$&
$3.090_{-0.050}^{+0.053}$&
$2.941_{-0.059}^{+0.057}$&
$2.678_{-0.054}^{+0.055}$&
$2.260_{-0.047}^{+0.046}$&
$1.648_{-0.038}^{+0.037}$&
$0.738_{-0.033}^{+0.035}$\\
\hline
\hline
$\Omega_\Lambda$&
$0.784_{-0.031}^{+0.030}$&
$0.787_{-0.031}^{+0.031}$&
$0.795_{-0.030}^{+0.031}$&
$0.795_{-0.030}^{+0.031}$&
$0.795_{-0.031}^{+0.031}$&
$0.791_{-0.032}^{+0.033}$\\
\hline
Age/Gyr&
$13.432_{-0.094}^{+0.092}$&
$13.418_{-0.097}^{+0.097}$&
$13.418_{-0.097}^{+0.097}$&
$13.415_{-0.098}^{+0.096}$&
$13.395_{-0.096}^{+0.094}$&
$13.400_{-0.091}^{+0.090}$\\
\hline
$\Omega_\mathrm{m}$&
$0.216_{-0.030}^{+0.031}$&
$0.213_{-0.031}^{+0.031}$&
$0.205_{-0.031}^{+0.030}$&
$0.205_{-0.031}^{+0.030}$&
$0.205_{-0.031}^{+0.031}$&
$0.209_{-0.033}^{+0.032}$\\
\hline
$\sigma_8$&
$0.808_{-0.049}^{+0.050}$&
$0.808_{-0.055}^{+0.054}$&
$0.796_{-0.047}^{+0.050}$&
$0.788_{-0.049}^{+0.049}$&
$0.787_{-0.050}^{+0.050}$&
$0.784_{-0.053}^{+0.052}$\\
\hline
$z_\mathrm{reion}$&
$15.4_{-2.0}^{+2.0}$&
$15.8_{-2.2}^{+2.3}$&
$15.8_{-2.3}^{+2.4}$&
$15.1_{-2.1}^{+2.1}$&
$14.7_{-1.9}^{+1.9}$&
$13.8_{-2.1}^{+2.1}$\\
\hline
$h$&
$0.770_{-0.031}^{+0.030}$&
$0.773_{-0.031}^{+0.032}$&
$0.781_{-0.032}^{+0.032}$&
$0.782_{-0.032}^{+0.032}$&
$0.782_{-0.033}^{+0.033}$&
$0.778_{-0.033}^{+0.034}$\\
\hline
$\Delta m^2$&
$0.07193_{-0.00072}^{+0.00077}$&
$0.1419_{-0.0015}^{+0.0015}$&
$0.2455_{-0.0022}^{+0.0022}$&
$0.3815_{-0.0028}^{+0.0027}$&
$0.5417_{-0.0029}^{+0.0028}$&
$0.7058_{-0.0031}^{+0.0033}$\\
\hline
\hline
$\chi^2$&
$11259$&
$11258$&
$11258$&
$11258$&
$11259$&
$11259$\\
\hline
$\Delta_\mathrm{AIC}$&
$8$&
$7$&
$7$&
$7$&
$8$&
$8$\\
\hline
\end{tabular}
\end{center}
\caption{\label{tab:wmap1step}
 1$\sigma$ constraints on the marginalised cosmological parameters for
 the $\Lambda$CDM step model using {\sl WMAP} data alone. The 6
 parameters in the upper part of the Table are varied by CosmoMC,
 while those in the lower part are derived quantities. The $\chi^2$ of
 the fit is given, as is the Akaike information criterion relative to
 the power-law $\Lambda$CDM model in Table~\ref{tab:tilt} (taking into
 account that $n$ is also a parameter).}
\end{table}

\begin{table}
\begin{center}
\begin{tabular}{|c|c|c|c|c|c|c|}
\hline 
$n$&
12&
13&
14&
15&
16&
17\\
\hline
\hline 
$\Omega_\mathrm{b} h^2$&
$0.02382_{-0.00040}^{+0.00042}$&
$0.02381_{-0.00039}^{+0.00042}$&
$0.02381_{-0.00043}^{+0.00040}$&
$0.02383_{-0.00038}^{+0.00039}$&
$0.02385_{-0.00041}^{+0.00041}$&
$0.02393_{-0.00040}^{+0.00039}$\\
\hline 
$\Omega_\mathrm{c} h^2$&
$0.1132_{-0.0053}^{+0.0055}$&
$0.1126_{-0.0056}^{+0.0058}$&
$0.1130_{-0.0056}^{+0.0055}$&
$0.1127_{-0.0054}^{+0.0057}$&
$0.1130_{-0.0055}^{+0.0055}$&
$0.1133_{+0.0055}^{+0.0054}$\\
\hline 
$\theta$&
$1.0460_{-0.0029}^{+0.0028}$&
$1.0458_{-0.0029}^{+0.0030}$&
$1.0459_{-0.0028}^{+0.0029}$&
$1.0459_{-0.0027}^{+0.0028}$&
$1.0461_{-0.0029}^{+0.0029}$&
$1.0469_{-0.0029}^{+0.0029}$\\
\hline 
$\tau$&
$0.138_{-0.013}^{+0.014}$&
$0.144_{-0.015}^{+0.016}$&
$0.138_{-0.016}^{+0.012}$&
$0.133_{-0.013}^{+0.013}$&
$0.128_{-0.012}^{+0.014}$&
$0.118_{-0.012}^{+0.014}$\\
\hline
$10^4 k_1/\mathrm{Mpc}^{-1}$&
$21.1_{-7.9}^{+4.6}$&
$14.1_{-6.6}^{+6.2}$&
$7.1_{-3.1}^{+1.0}$&
$4.9_{-1.9}^{+2.0}$&
$3.5_{-1.2}^{+0.4}$&
$2.5_{-1.1}^{+0.9}$\\
\hline
$\ln\left(10^{10}\mathcal{P_{R}}^{(0)}\right)$&
$3.107_{-0.049}^{+0.051}$&
$2.959_{-0.051}^{+0.054}$&
$2.691_{-0.048}^{+0.051}$&
$2.280_{-0.041}^{+0.043}$&
$1.667_{-0.039}^{+0.039}$&
$0.755_{-0.029}^{+0.032}$\\
\hline
\hline
$\Omega_\Lambda$&
$0.746_{-0.022}^{+0.023}$&
$0.747_{-0.023}^{+0.023}$&
$0.745_{-0.023}^{+0.023}$&
$0.747_{-0.022}^{+0.023}$&
$0.746_{-0.023}^{+0.023}$&
$0.746_{-0.022}^{+0.022}$\\
\hline
Age/Gyr&
$13.443_{-0.092}^{+0.094}$&
$13.446_{-0.096}^{+0.095}$&
$13.443_{-0.095}^{+0.094}$&
$13.441_{-0.087}^{+0.092}$&
$13.435_{-0.094}^{+0.094}$&
$13.407_{-0.095}^{+0.092}$\\
\hline
$\Omega_\mathrm{m}$&
$0.255_{-0.023}^{+0.024}$&
$0.253_{-0.023}^{+0.024}$&
$0.255_{-0.023}^{+0.023}$&
$0.253_{-0.023}^{+0.022}$&
$0.254_{-0.023}^{+0.023}$&
$0.254_{-0.022}^{+0.022}$\\
\hline
$\sigma_8$&
$0.863_{-0.033}^{+0.034}$&
$0.864_{-0.035}^{+0.035}$&
$0.862_{-0.034}^{+0.033}$&
$0.856_{-0.033}^{+0.034}$&
$0.853_{-0.033}^{+0.034}$&
$0.849_{-0.034}^{+0.034}$\\
\hline
$z_\mathrm{reion}$&
$14.8_{-2.0}^{+2.1}$&
$15.2_{-2.2}^{+2.3}$&
$14.8_{-2.2}^{+2.2}$&
$14.4_{-2.0}^{+2.1}$&
$14.0_{-2.1}^{+2.1}$&
$13.3_{-2.0}^{+2.1}$\\
\hline
$h$&
$0.734_{-0.019}^{+0.019}$&
$0.735_{-0.020}^{+0.020}$&
$0.734_{-0.019}^{+0.020}$&
$0.736_{-0.019}^{+0.020}$&
$0.735_{-0.020}^{+0.019}$&
$0.737_{-0.018}^{+0.019}$\\
\hline
$\Delta m^2$&
$0.07219_{-0.00070}^{+0.00073}$&
$0.1424_{-0.0013}^{+0.0014}$&
$0.2460_{-0.0020}^{+0.0021}$&
$0.3826_{-0.0024}^{+0.0025}$&
$0.5431_{-0.0030}^{+0.0030}$&
$0.7074_{-0.0028}^{+0.0030}$\\
\hline
\hline
$\chi^2$&
$11278$&
$11278$&
$11277$&
$11278$&
$11278$&
$11278$\\
\hline
$\Delta_\mathrm{AIC}$&
$9$&
$9$&
$8$&
$9$&
$9$&
$9$\\
\hline
\end{tabular}
\end{center}
\caption{\label{tab:pow1step}
 1$\sigma$ constraints on the marginalised cosmological parameters for
 the $\Lambda$CDM step model using {\sl WMAP} + {\sl SDSS} data. The 6
 parameters in the upper part of the Table are varied by CosmoMC,
 while those in the lower part are derived quantities. The $\chi^2$ of
 the fit is given, as is the Akaike information criterion relative to
 the power-law $\Lambda$CDM model in Table~\ref{tab:tilt} (taking into
 account that $n$ is also a parameter).}
\end{table}

\begin{table}
\begin{center}
\begin{tabular}{|c|c|c|c|c|c|c|}
\hline 
$n$&
12&
13&
14&
15&
16&
17\\
\hline
\hline 
$\Omega_\mathrm{b} h^2$&
$0.02383_{-0.00039}^{+0.00039}$&
$0.02384_{-0.00040}^{+0.00039}$&
$0.02386_{-0.00041}^{+0.00040}$&
$0.02388_{-0.00040}^{+0.00041}$&
$0.02389_{-0.00040}^{+0.00041}$&
$0.02393_{-0.00040}^{+0.00039}$\\
\hline 
$\Omega_\mathrm{c} h^2$&
$0.1160_{-0.0044}^{+0.0044}$&
$0.1159_{-0.0046}^{+0.0045}$&
$0.1159_{-0.0045}^{+0.0046}$&
$0.1159_{-0.0047}^{+0.0045}$&
$0.1159_{-0.0044}^{+0.0047}$&
$0.1163_{-0.0044}^{+0.0044}$\\
\hline 
$\theta$&
$1.0460_{-0.0029}^{+0.0029}$&
$1.0460_{-0.0028}^{+0.0028}$&
$1.0461_{-0.0027}^{+0.0027}$&
$1.0464_{-0.0028}^{+0.0030}$&
$1.0465_{-0.0028}^{+0.0029}$&
$1.0468_{-0.0028}^{+0.0028}$\\
\hline 
$\tau$&
$0.133_{-0.013}^{+0.013}$&
$0.137_{-0.014}^{+0.013}$&
$0.128_{-0.013}^{+0.012}$&
$0.125_{-0.012}^{+0.013}$&
$0.120_{-0.011}^{+0.013}$&
$0.113_{-0.012}^{+0.014}$\\
\hline
$10^4 k_1/\mathrm{Mpc}^{-1}$&
$20.2_{-8.3}^{+4.8}$&
$13.1_{-6.5}^{+6.4}$&
$6.3_{-2.2}^{+1.5}$&
$4.8_{-1.8}^{+1.8}$&
$3.5_{-1.0}^{+0.4}$&
$2.5_{-1.1}^{+1.0}$\\
\hline
$\ln\left(10^{10}\mathcal{P_R}^{(0)}\right)$&
$3.108_{-0.048}^{+0.050}$&
$2.957_{-0.051}^{+0.053}$&
$2.683_{-0.047}^{+0.047}$&
$2.278_{-0.045}^{+0.045}$&
$1.663_{-0.038}^{+0.038}$&
$0.756_{-0.032}^{+0.032}$\\
\hline
$b_\mathrm{LRG}$&
$2.104_{-0.092}^{+0.094}$&
$2.100_{-0.092}^{+0.094}$&
$2.117_{-0.089}^{+0.090}$&
$2.119_{-0.091}^{+0.091}$&
$2.135_{-0.092}^{+0.092}$&
$2.143_{-0.091}^{+0.093}$\\
\hline
\hline
$\Omega_\Lambda$&
$0.732_{-0.018}^{+0.018}$&
$0.732_{-0.018}^{+0.018}$&
$0.733_{-0.020}^{+0.019}$&
$0.734_{-0.019}^{+0.019}$&
$0.734_{-0.019}^{+0.018}$&
$0.733_{-0.019}^{+0.018}$\\
\hline
Age/Gyr&
$13.456_{-0.095}^{+0.094}$&
$13.454_{-0.090}^{+0.093}$&
$13.450_{-0.093}^{+0.088}$&
$13.440_{-0.098}^{+0.096}$&
$13.436_{-0.094}^{+0.095}$&
$13.425_{-0.093}^{+0.095}$\\
\hline
$\Omega_\mathrm{m}$&
$0.268_{-0.018}^{+0.018}$&
$0.267_{-0.018}^{+0.018}$&
$0.267_{-0.019}^{+0.020}$&
$0.266_{-0.019}^{+0.019}$&
$0.266_{-0.018}^{+0.019}$&
$0.267_{-0.019}^{+0.019}$\\
\hline
$\sigma_8$&
$0.878_{-0.032}^{+0.032}$&
$0.878_{-0.032}^{+0.032}$&
$0.870_{-0.031}^{+0.031}$&
$0.869_{-0.031}^{+0.031}$&
$0.864_{-0.030}^{+0.031}$&
$0.861_{-0.031}^{+0.031}$\\
\hline
$z_\mathrm{reion}$&
$14.5_{-2.0}^{+2.1}$&
$14.8_{-2.2}^{+2.3}$&
$14.1_{-2.1}^{+2.1}$&
$13.9_{-2.1}^{+2.2}$&
$13.4_{-2.1}^{+2.1}$&
$12.9_{-2.2}^{+2.2}$\\
\hline
$h$&
$0.708_{-0.015}^{+0.015}$&
$0.724_{-0.015}^{+0.015}$&
$0.724_{-0.016}^{+0.016}$&
$0.725_{-0.015}^{+0.015}$&
$0.726_{-0.015}^{+0.015}$&
$0.725_{-0.015}^{+0.015}$\\
\hline
$\Delta m^2$&
$0.07152_{-0.00069}^{+0.00072}$&
$0.1423_{-0.0013}^{+0.0014}$&
$0.2457_{-0.0019}^{+0.0019}$&
$0.3825_{-0.0027}^{+0.0027}$&
$0.5428_{-0.0029}^{+0.0029}$&
$0.7075_{-0.0031}^{+0.0030}$\\
\hline
\hline
$\chi^2$&
$11282$&
$11282$&
$11282$&
$11282$&
$11282$&
$11283$\\
\hline
$\Delta_\mathrm{AIC}$&
$9$&
$9$&
$9$&
$9$&
$9$&
$10$\\
\hline
\end{tabular}
\end{center}
\caption{\label{tab:corr1step}
 1$\sigma$ constraints on the marginalised cosmological parameters for
 the $\Lambda$CDM step model using {\sl WMAP} + LRG data. The 7
 parameters in the upper part of the Table are varied by CosmoMC,
 while those in the lower part are derived quantities. The $\chi^2$ of
 the fit is given, as is the Akaike information criterion relative to
 the power-law $\Lambda$CDM model in Table~\ref{tab:tilt} (taking into
 account that $n$ is also a parameter).}
\end{table}

\begin{table}
\begin{center}
\begin{tabular}{|c|c|c|c|c|c|c|}
\hline 
$n$&
12&
13&
14&
15&
16&
17\\
\hline
\hline 
$\Omega_\mathrm{b} h^2$&
$0.02386_{-0.00040}^{+0.00040}$&
$0.02384_{-0.00039}^{+0.00039}$&
$0.02385_{-0.00041}^{+0.00040}$&
$0.02387_{-0.00040}^{+0.00040}$&
$0.02391_{-0.00042}^{+0.00041}$&
$0.02395_{-0.00040}^{+0.00039}$\\
\hline 
$\Omega_\mathrm{c} h^2$&
$0.1170_{-0.0040}^{+0.0039}$&
$0.1172_{-0.0041}^{+0.0041}$&
$0.1172_{-0.0040}^{+0.0040}$&
$0.1171_{-0.0040}^{+0.0040}$&
$0.1176_{-0.0043}^{+0.0040}$&
$0.1174_{-0.0039}^{+0.0037}$\\
\hline 
$\theta$&
$1.0460_{-0.0028}^{+0.0028}$&
$1.0461_{-0.0028}^{+0.0028}$&         
$1.0463_{-0.0028}^{+0.0028}$&
$1.0467_{-0.0027}^{-0.0027}$&
$1.0466_{-0.0027}^{+0.0027}$&
$1.0469_{-0.0025}^{+0.0027}$\\
\hline 
$\tau$&
$0.130_{-0.012}^{+0.013}$&
$0.135_{-0.014}^{+0.015}$&
$0.126_{-0.014}^{+0.013}$&
$0.124_{-0.012}^{+0.013}$&
$0.118_{-0.011}^{+0.013}$&
$0.111_{-0.012}^{+0.014}$\\
\hline
$10^4 k_1/\mathrm{Mpc}^{-1}$&
$19.1_{-9.3}^{+4.6}$&
$13.1_{-6.4}^{+6.2}$&
$6.5_{-2.6}^{+1.4}$&
$4.4_{-1.6}^{+1.5}$&
$3.2_{-1.0}^{+0.6}$&
$2.5_{-1.1}^{+1.0}$\\
\hline
$\ln\left(10^{10}\mathcal{P_R}^{(0)}\right)$&
$3.106_{-0.047}^{+0.049}$&
$2.960_{-0.051}^{+0.051}$&
$2.685_{-0.047}^{+0.047}$&
$2.279_{-0.040}^{+0.040}$&
$1.667_{-0.037}^{+0.037}$&
$0.757_{-0.032}^{+0.031}$\\
\hline
$b_\mathrm{LRG}$&
$2.110_{-0.089}^{+0.090}$&
$2.098_{-0.091}^{+0.091}$&
$2.118_{-0.094}^{+0.094}$&
$2.122_{-0.088}^{+0.089}$&
$2.133_{-0.091}^{+0.088}$&
$2.143_{-0.092}^{+0.092}$\\
\hline
\hline
$\Omega_\Lambda$&
$0.727_{-0.016}^{+0.016}$&
$0.727_{-0.017}^{+0.017}$&
$0.727_{-0.017}^{+0.017}$&
$0.728_{-0.016}^{+0.016}$&
$0.726_{-0.017}^{+0.017}$&
$0.728_{-0.015}^{+0.016}$\\
\hline
Age/Gyr&
$13.456_{-0.093}^{+0.090}$&
$13.457_{-0.092}^{+0.093}$&
$13.451_{-0.092}^{+0.092}$&
$13.446_{-0.090}^{+0.091}$&
$13.439_{-0.096}^{+0.093}$&
$13.426_{-0.090}^{+0.089}$\\
\hline
$\Omega_\mathrm{m}$&
$0.273_{-0.016}^{+0.016}$&
$0.273_{-0.017}^{+0.017}$&
$0.273_{-0.017}^{+0.017}$&
$0.272_{-0.016}^{+0.016}$&
$0.274_{-0.017}^{+0.017}$&
$0.272_{-0.016}^{+0.015}$\\
\hline
$\sigma_8$&
$0.879_{-0.028}^{+0.029}$&
$0.884_{-0.029}^{+0.029}$&
$0.877_{-0.029}^{+0.029}$&
$0.874_{-0.028}^{+0.028}$&
$0.873_{-0.030}^{+0.028}$&
$0.866_{-0.029}^{+0.028}$\\
\hline
$z_\mathrm{reion}$&
$14.3_{-2.0}^{+2.1}$&
$14.7_{-2.2}^{+2.2}$&
$14.0_{-2.2}^{+2.2}$&
$13.8_{-2.0}^{+2.0}$&
$13.4_{-2.1}^{+2.1}$&
$12.8_{-2.2}^{+2.2}$\\
\hline
$h$&
$0.719_{-0.014}^{+0.013}$&
$0.719_{-0.013}^{+0.014}$&
$0.720_{-0.014}^{+0.014}$&
$0.720_{-0.013}^{+0.014}$&
$0.720_{-0.013}^{+0.013}$&
$0.722_{-0.013}^{+0.013}$\\
\hline
$\Delta m^2$&
$0.07219_{-0.00068}^{+0.00071}$&
$0.1424_{-0.0013}^{+0.0013}$&
$0.2458_{-0.0020}^{+0.0020}$&
$0.3826_{-0.0024}^{+0.0023}$&
$0.5431_{-0.0029}^{+0.0029}$&
$0.7076_{-0.0030}^{+0.0030}$\\
\hline
\hline
$\chi^{2}$&
$11299$&
$11299$&
$11299$&
$11299$&
$11299$&
$11299$\\
\hline
$\Delta_\mathrm{AIC}$&
$9$&
$9$&
$9$&
$9$&
$9$&
$9$\\
\hline
\end{tabular}
\end{center}
\caption{\label{tab:all1step}
 1$\sigma$ constraints on the marginalised cosmological parameters for
 the $\Lambda$CDM step model using {\sl WMAP} + {\sl SDSS} + LRG
 data. The 7 parameters in the upper part of the Table are varied by
 CosmoMC, while those in the lower part are derived quantities. The
 $\chi^2$ of the fit is given, as is the Akaike information criterion
 relative to the power-law $\Lambda$CDM model in Table~\ref{tab:tilt}
 (taking into account that $n$ is also a parameter).}
\end{table}

\begin{figure}[tbh]
\includegraphics[angle=-90,scale=0.5]{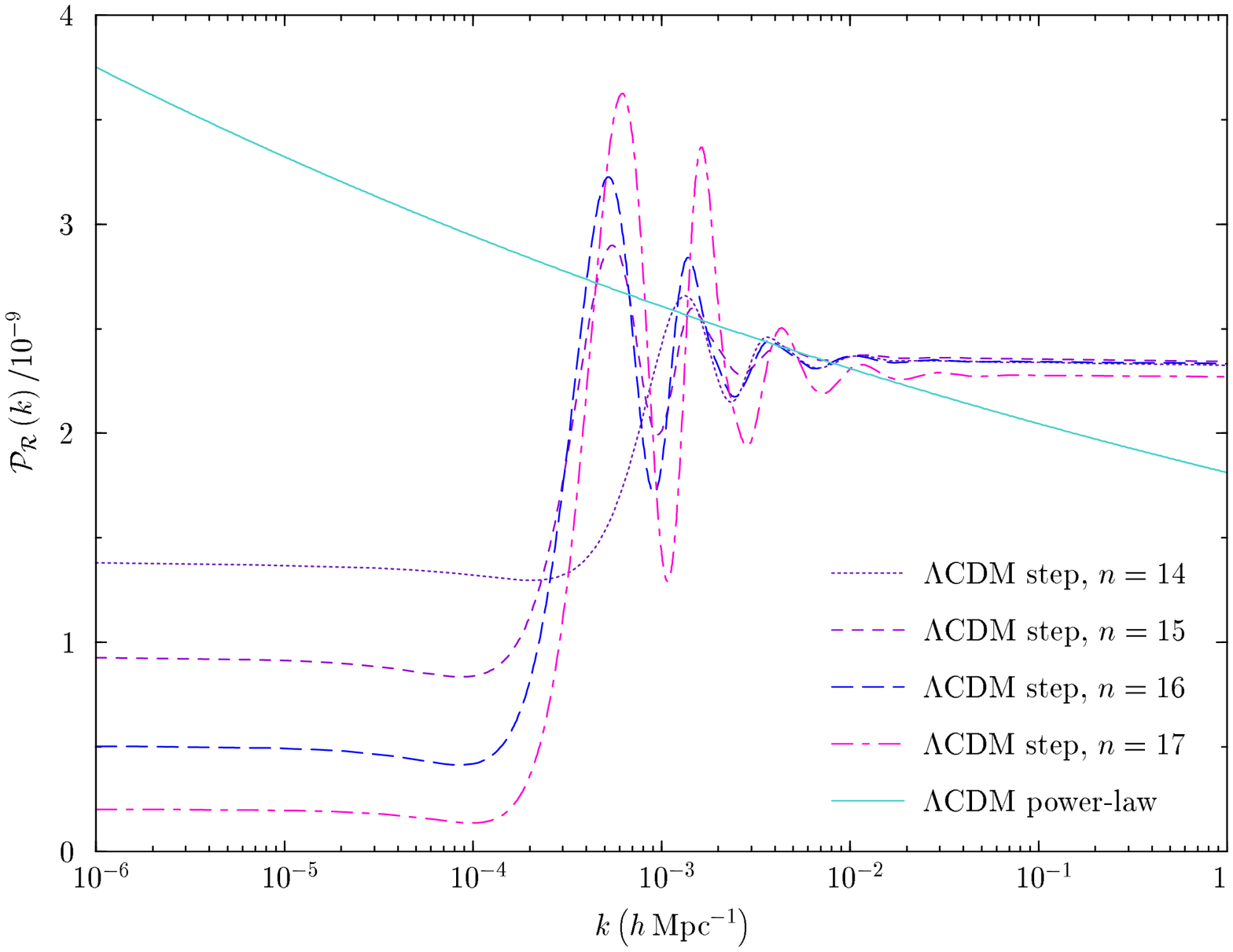}
\caption{\label{allprimspec3a} 
  The best-fit primordial perturbation spectra for the $\Lambda$CDM
  step model. Note the suppression of power at the wavenumber
  coresponding to the present Hubble radius: $H_0 \simeq 3 \times
  10^{-4}\,h\,\mathrm{Mpc^{-1}}$.}
\end{figure}

To understand this, recall that the amplitude of the primordial
perturbation spectrum is well constrained by the TT spectrum on medium
scales, but is uncertain on large scales due to cosmic variance.
Therefore the data can accommodate a large step in $\mathcal{P_R}$
provided that the top of the step is at the right height to match the
TT spectrum. Consequently $\mathcal{P_R}^{(0)}$ must vary with the
inflaton mass change in such a way as to leave the primordial
perturbation spectrum invariant on smaller scales. This is seen in
Fig.\ref{allprimspec3a} which shows the primordial spectra for the
best-fit models.

\begin{figure}[tbh]
\includegraphics[angle=-90,scale=0.5]{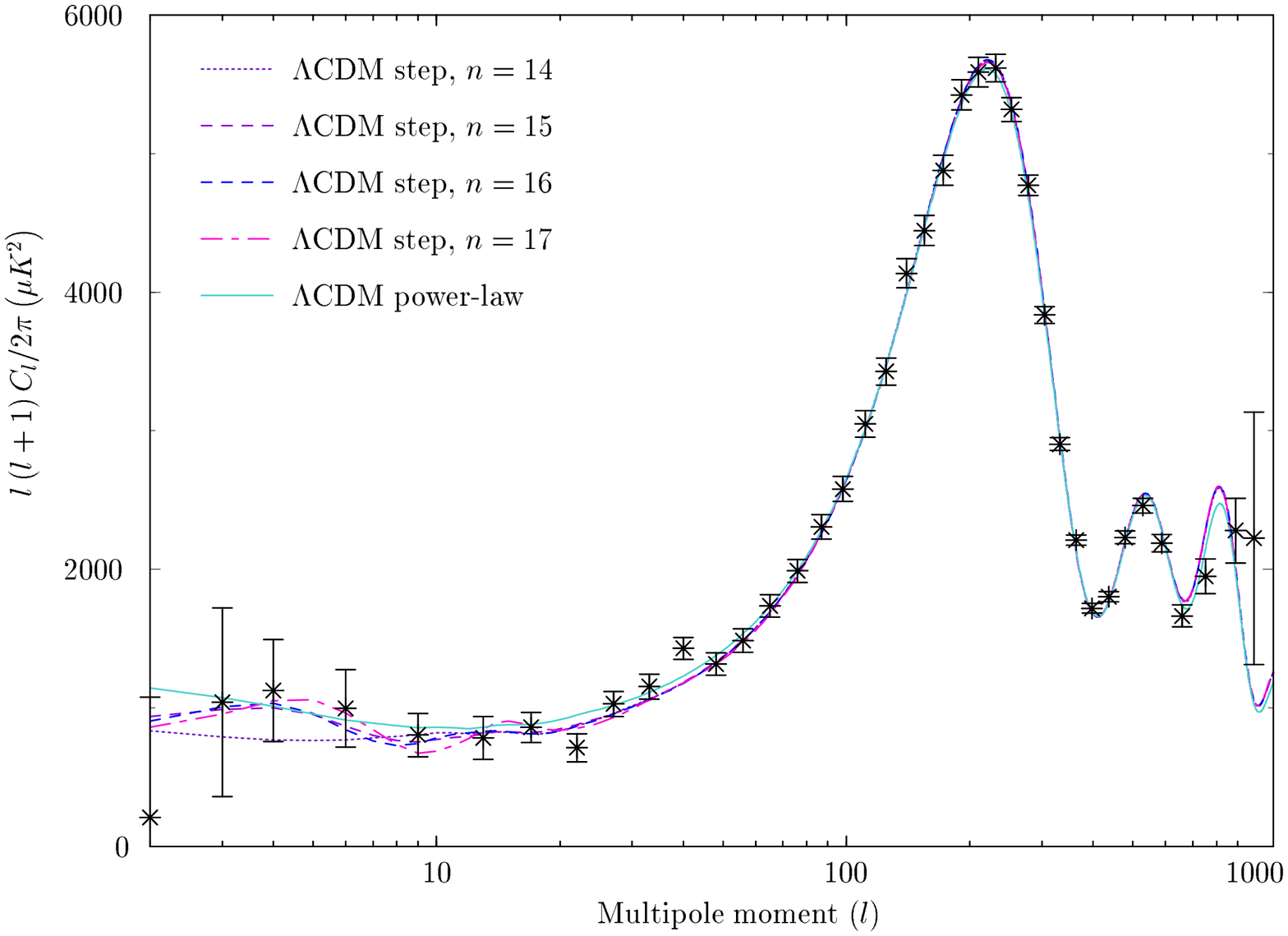}
\caption{\label{allcltt3a} 
 The best-fit TT spectra for the $\Lambda$CDM step model, with
 {\sl WMAP} data.}
\end{figure}

\begin{figure}[tbh]
\includegraphics[angle=-90,scale=0.5]{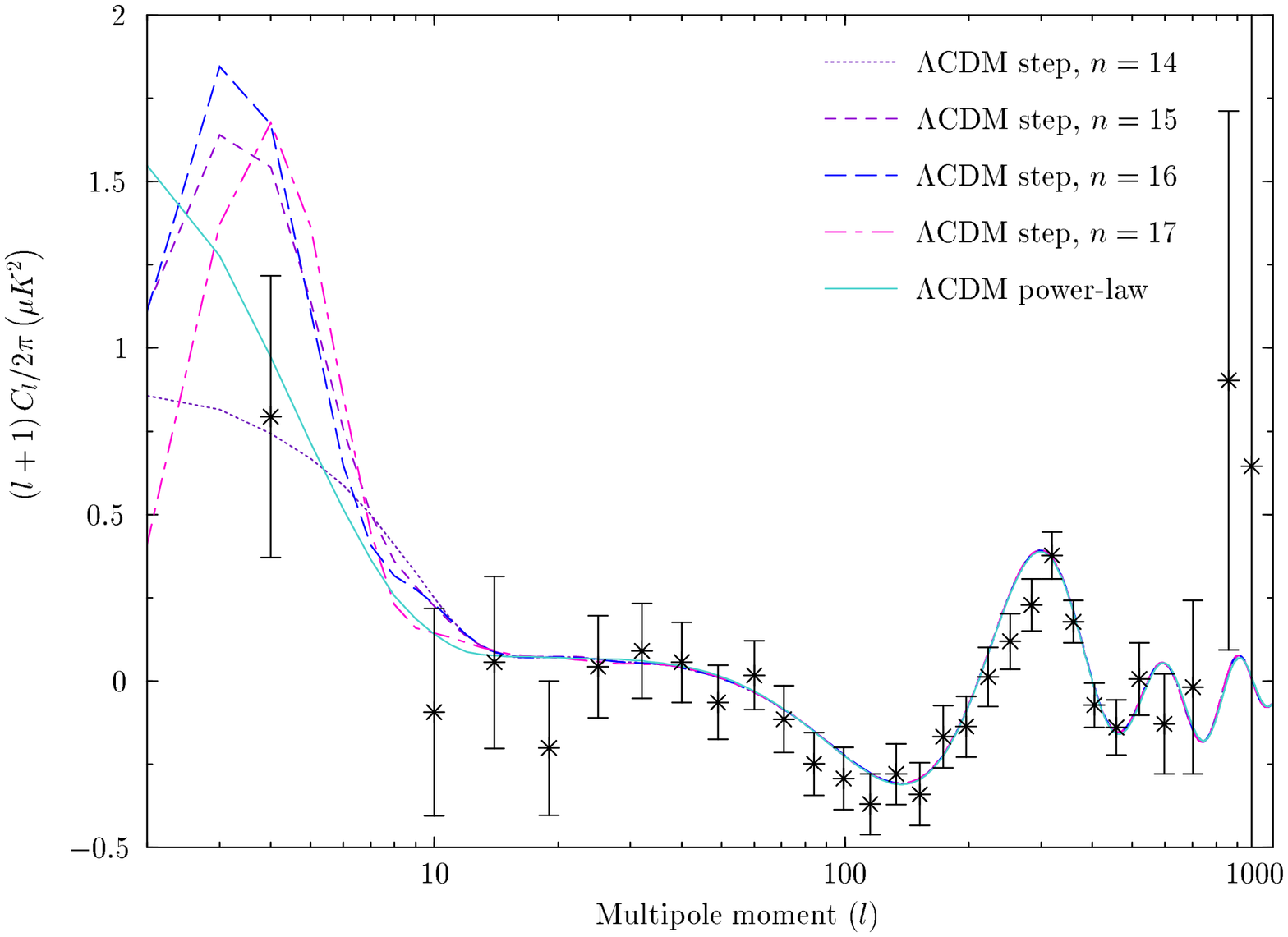}
\caption{\label{allclte2a} 
 The best-fit TE spectra for the $\Lambda$CDM step model, with
 {\sl WMAP} data.}
\end{figure}

The position of the step in $\mathcal{P_R}(k)$ for $n=14$ means that
the associated TT and TE spectra are lower on large scales than
those of the best-fit power-law model --- see Figs.\ref{allcltt3a} and
\ref{allclte2a}. For $n=15$ to $n=17$ the lower plateau of the step is
too far outside the Hubble radius to suppress the TT and TE spectra.
However the primordial spectrum has a prominent first peak at $k
\simeq 6\times 10^{-4}$ $h$ Mpc$^{-1}$. This introduces a maximum in
the TT spectrum centred on $\ell=4$ which fits the excess power seen
there by {\sl WMAP}. There is a corresponding peak at low $\ell$ in
the TE spectrum. The peak at $k \simeq 1.5 \times
10^{-3}\,h\,\mathrm{Mpc^{-1}}$ in the primordial spectrum (for $n=14$
to $n=17$) increases the amplitude of the TT spectrum around
$\ell=15$. However the glitches in the TT spectrum at $\ell=22$ and
$\ell=40$ are too sharp to match the oscillations produced by the
mechanism considered here.

\begin{figure}[tbh]
\includegraphics[angle=-90,scale=0.5]{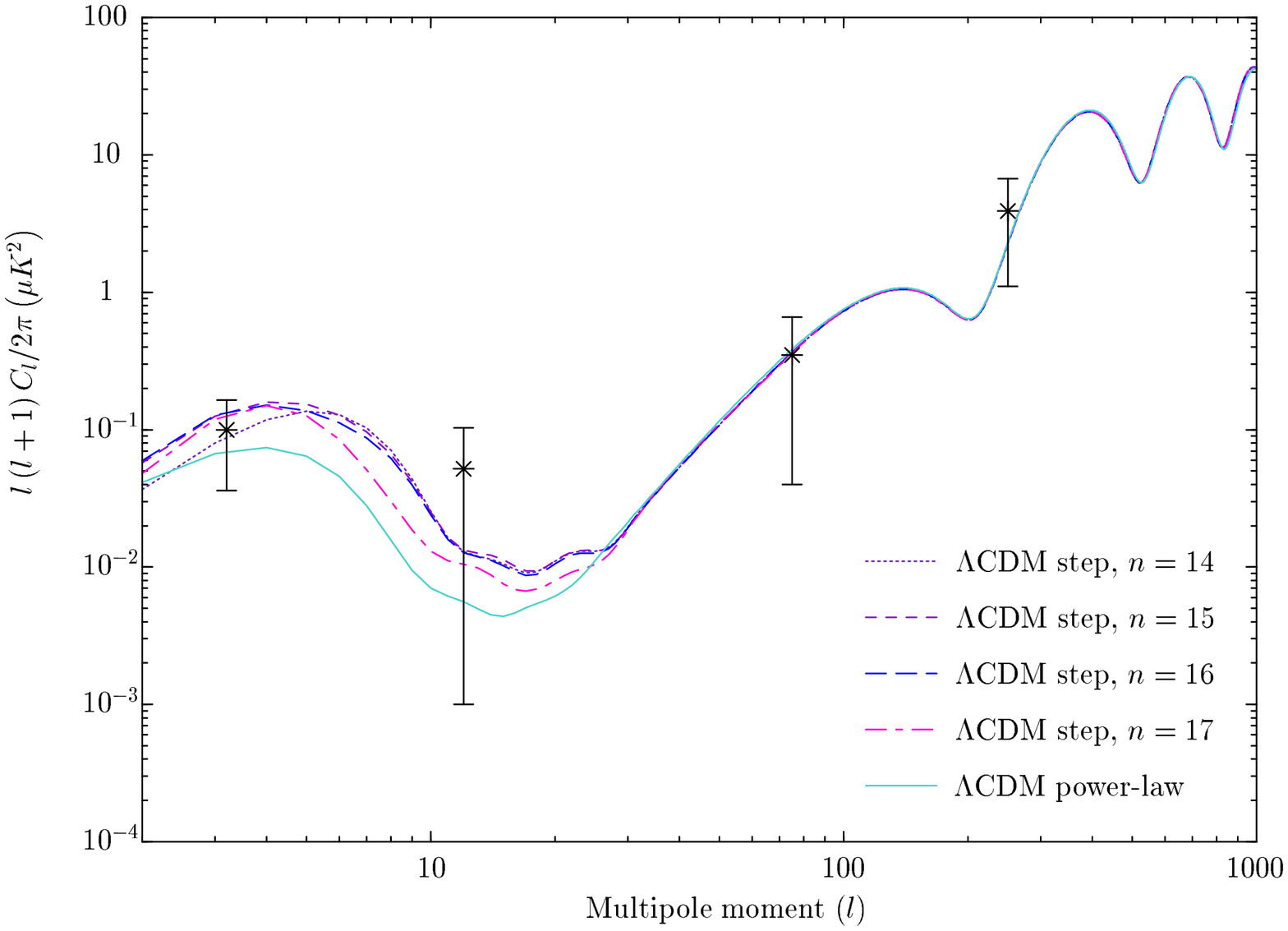}
\caption{\label{allclee1a} 
 The best-fit EE spectra for the $\Lambda$CDM step model, with
 {\sl WMAP} data.}
\end{figure}       
 
\begin{figure}[tbh]
\includegraphics[angle=-90,scale=0.5]{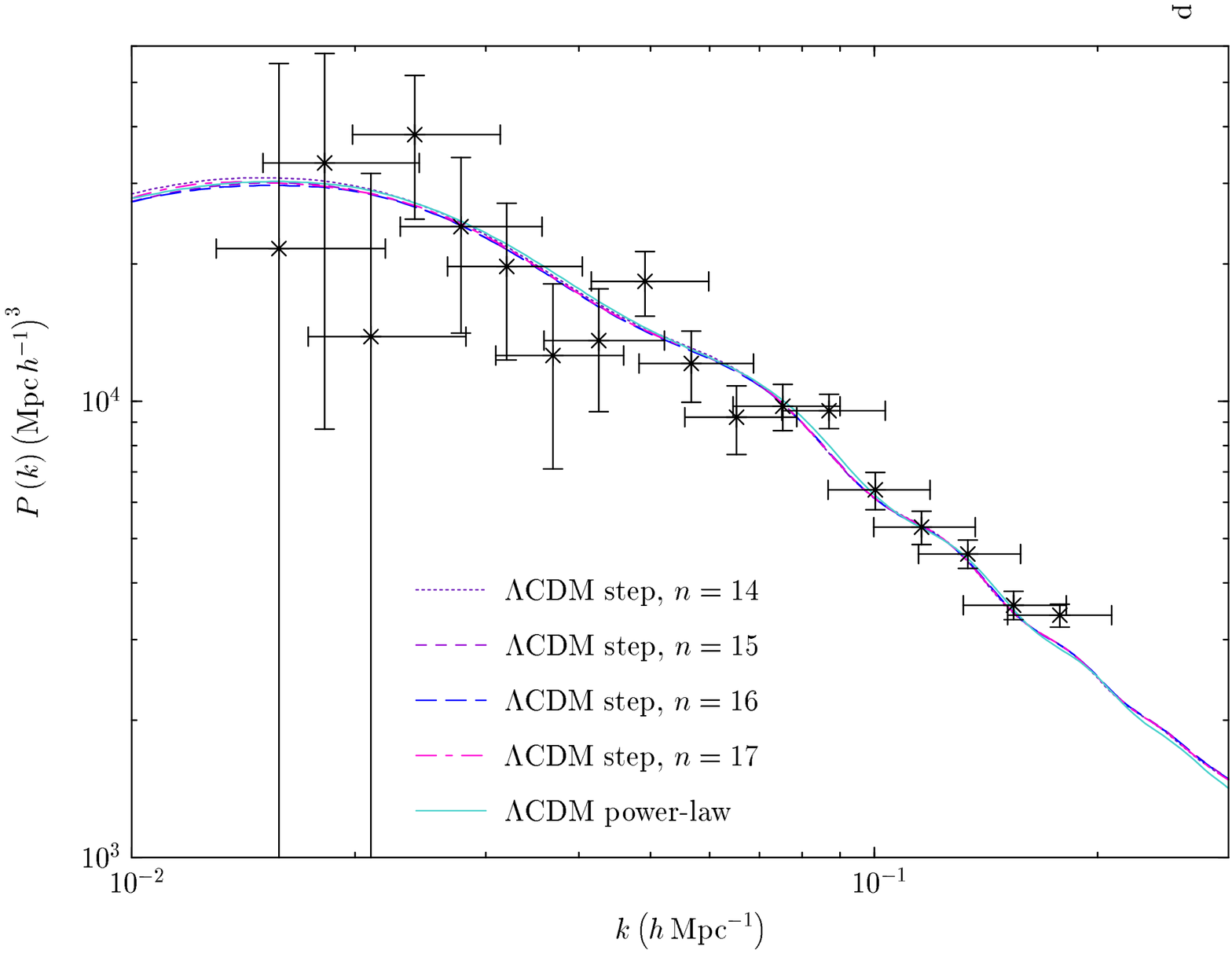}
\caption{\label{allmatpow3a} The best-fit matter power spectra for the
$\Lambda$CDM step model, with {\sl SDSS} data.}
\end{figure}

\begin{figure}[tbh]
\includegraphics[angle=-90,scale=0.5]{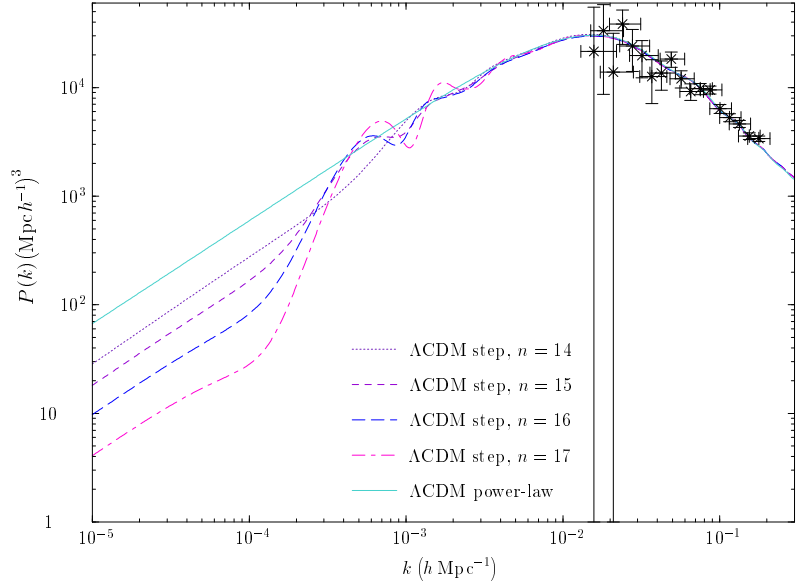}
\caption{\label{allmatpow3c} 
 Oscillations in the matter power spectra on very large scales in the
 $\Lambda$CDM step model.}
\end{figure}

\begin{figure}[tbh]
\includegraphics[angle=-90,scale=0.5]{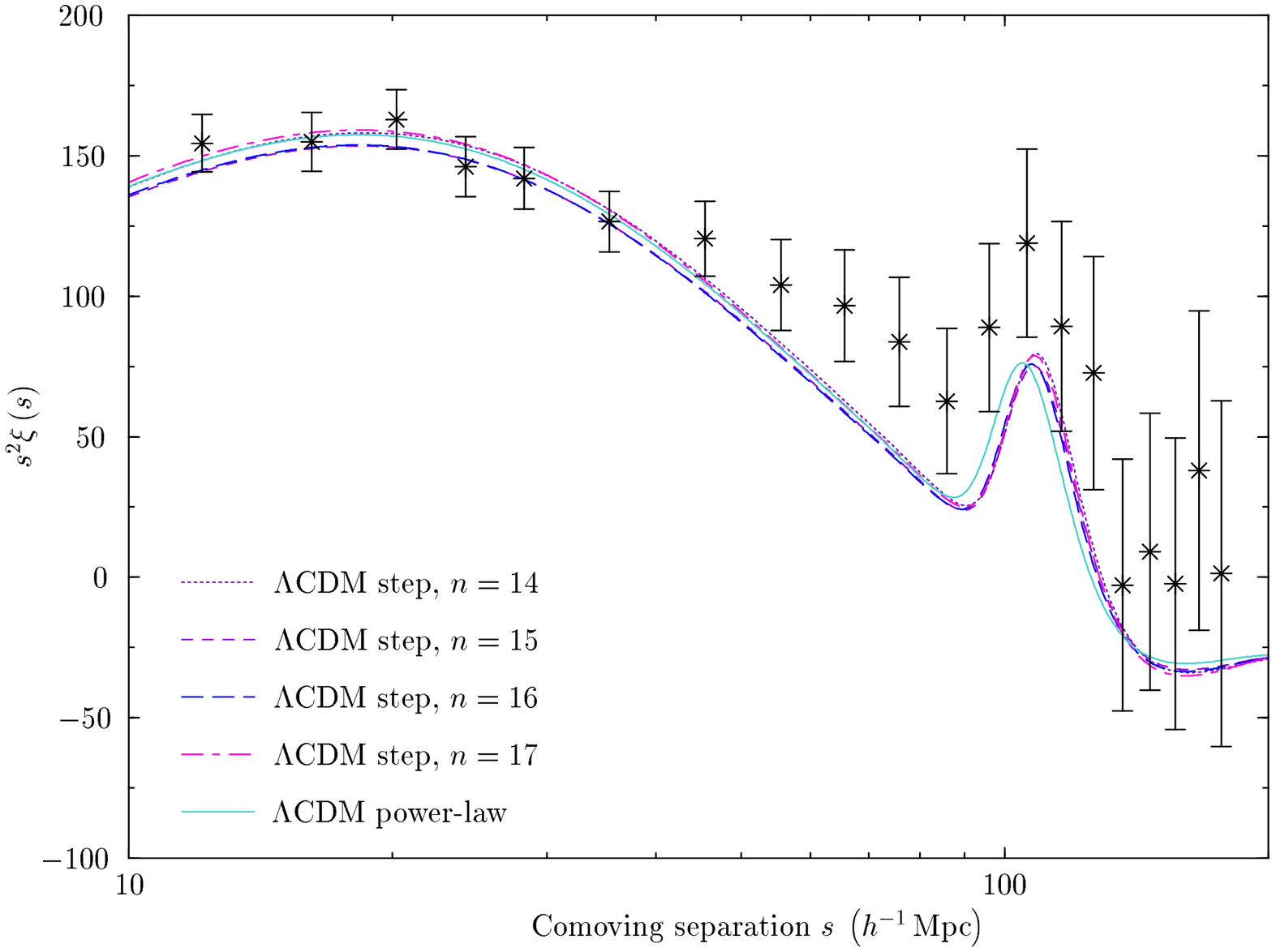}
\caption{\label{allcorr3a} The best-fit two-point galaxy correlation
  functions in the $\Lambda$CDM step model, with LRG data; the spatial
  scales have been shifted using eq.(\ref{scale}) to match the same
  $\Lambda$CDM cosmology and enable comparison. }
\end{figure}

As seen in Fig.\ref{allmatpow3a}, the {\sl SDSS} galaxy power spectrum
is well matched in all cases but the features in the spectrum due to
the phase transition appear far outside the scales probed by redshift
surveys (see Fig.\ref{allmatpow3c}). Moreover a step in the matter
power spectrum does not significantly alter the two-point correlation
function in a $\Lambda$CDM universe as shown in
Fig.\ref{allcorr3a}. Just as for the concordance power-law
$\Lambda$CDM model however, the amplitude of the predicted BAO peak is
too low by a factor of $\sim 2$, although its predicted position
does match the data \cite{Blanchard:2005ev}. 

We conclude therefore that the observations favour a `tilted'
primordial spectrum over a scale-invariant one, as has been emphasised
already by the {\sl WMAP} team \cite{Spergel:2006hy}. However the
spectrum is not required to be {\em scale-free}. Very different
primordial spectra (as shown in Fig.\ref{allprimspec3a}) provide just
as good a fit to the data, although this is admittedly penalised by
the Akaike information criterion taking into account the 2 additional
parameters characterising the step in the spectrum. This motivates us
to ask whether even the best-fit cosmological model might be altered
if a more radical departure from a scale-free spectrum is considered.
For example in multiple inflation \cite{Adams:1997de}, two phase
transitions can occur in rapid succession resulting in a `bump' in the
primordial spectrum. Such a feature has been advocated earlier on
empirical grounds for fitting a $\Lambda$CDM model \cite{Silk:2000sn}.

\subsection{CHDM `bump' model}  

\begin{figure}[tbh]
\includegraphics[angle=-90,scale=0.5]{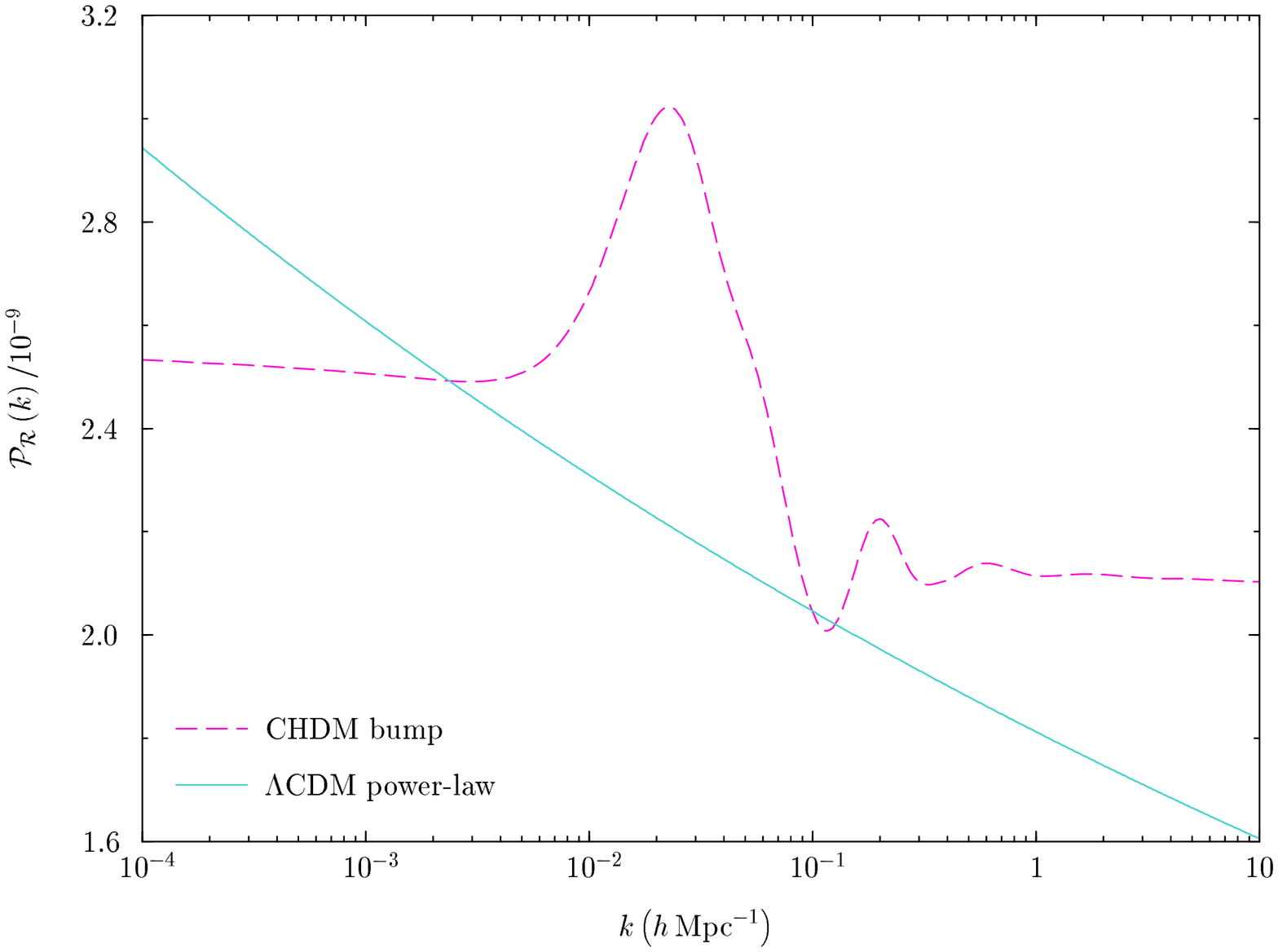}
\caption{\label{allprimspec3b}
 The primordial perturbation spectrum for the CHDM bump model with
 $n_1=12$ and $n_2=13$, compared to the $\Lambda$CDM power-law model
 with $n_\mathrm{s} \simeq 0.95$.}
\end{figure}

\begin{figure}[tbh]
\includegraphics[angle=-90,scale=0.5]{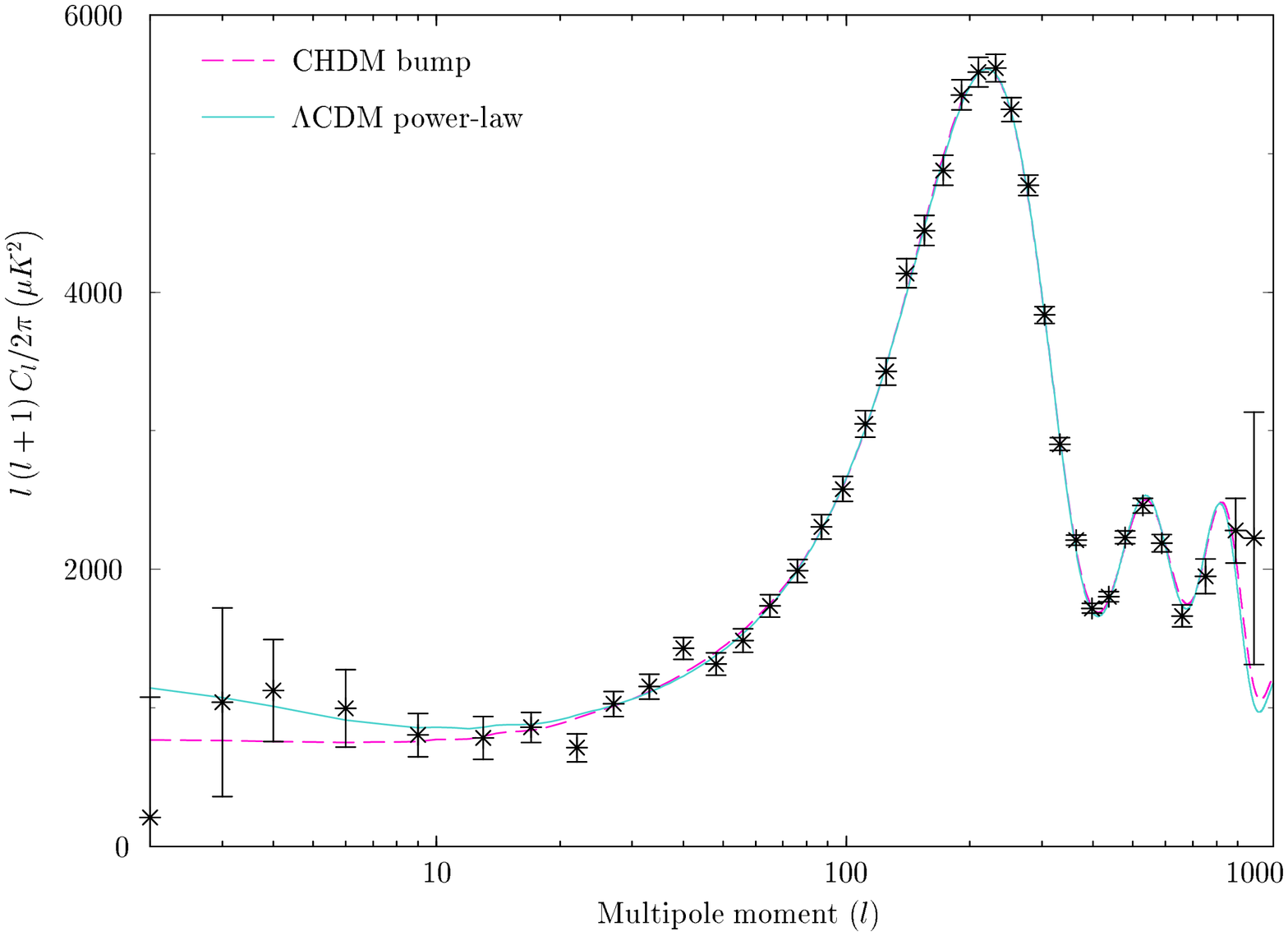}
\caption{\label{allcltt3b} The best-fit TT spectrum
for the CHDM bump model, with {\sl WMAP} data.}
\end{figure}

\begin{figure}[tbh]
\includegraphics[angle=-90,scale=0.5]{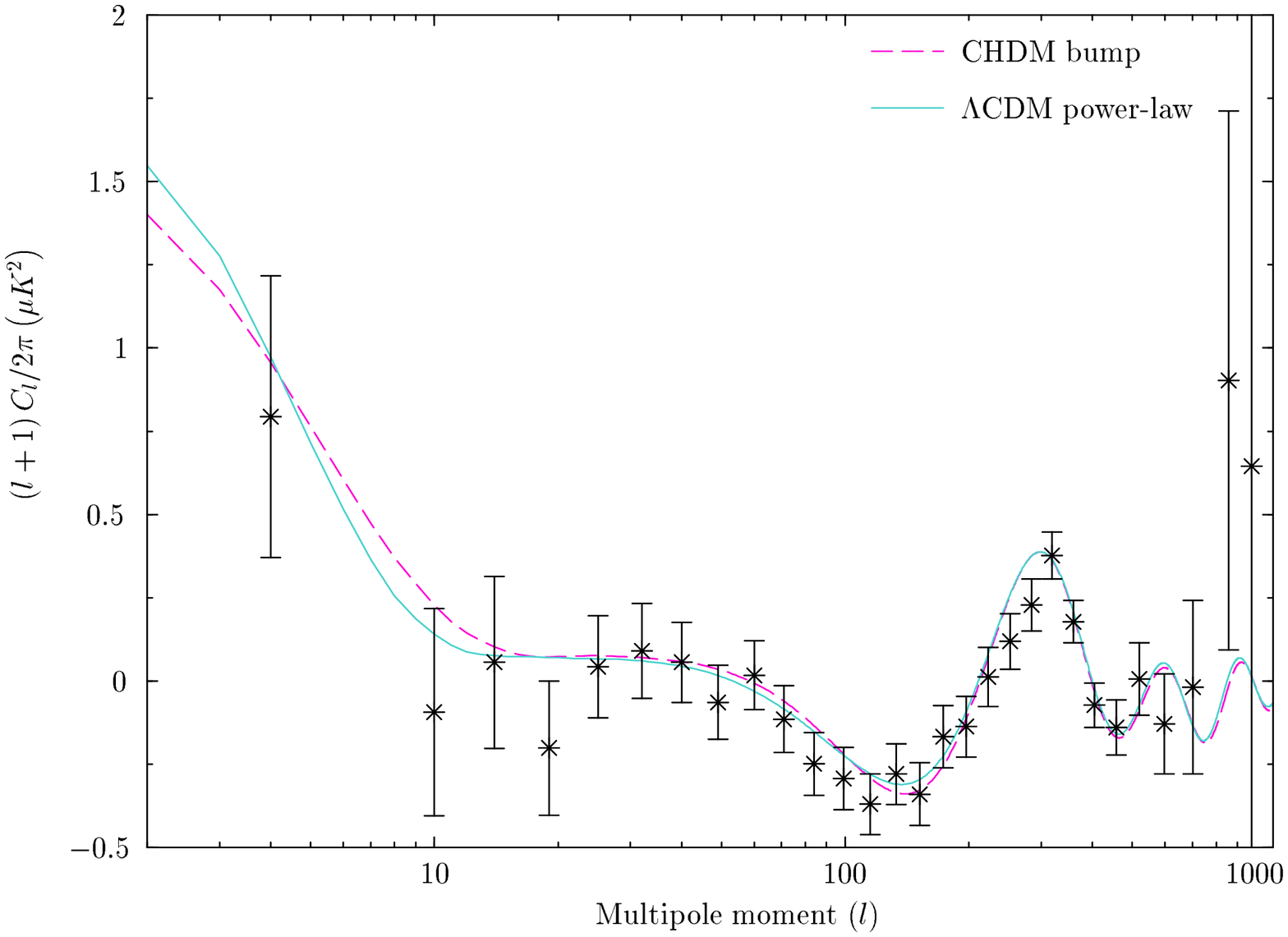}
\caption{\label{allclte3b} The best fit TE spectrum for the
CHDM bump model, with {\sl WMAP} data.}
\end{figure}

\begin{figure}[tbh]
\includegraphics[angle=-90,scale=0.5]{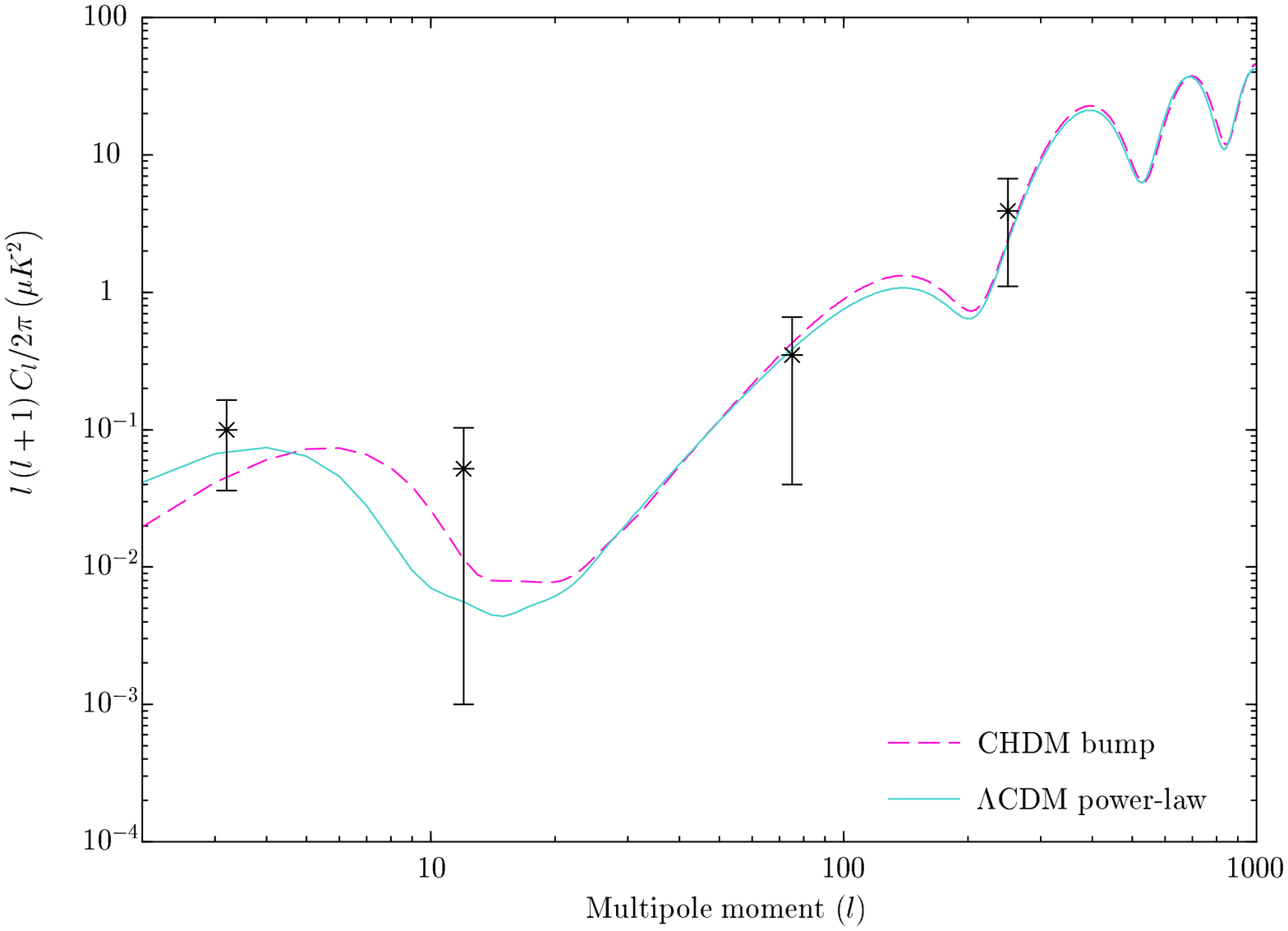}
\caption{\label{allclee1b} 
 The best fit EE spectrum for the CHDM bump model, with {\sl WMAP}
 data.}
\end{figure}

We consider now a primordial spectrum with a bump at $k \simeq 2
\times 10^{-3}\,h\,\mathrm{Mpc^{-1}}$ generated by 2 successive phase
transitions which cause an upward step followed by a slightly larger
downward step in the amplitude of the primordial perturbation, as
shown in Fig.\ref{allprimspec3b}. This boosts the amplitude of the TT
spectrum on the left of the first acoustic peak but suppresses the
second and third peaks. This is all that is necessary to fit the {\sl
WMAP} data to an Einstein-de Sitter cosmology as seen in
Fig.\ref{allcltt3b}. Since there is no late ISW effect, the amplitude
is smaller on large scales than in an universe with dark energy. The
fits to the TE and EE spectra are also good as shown in
Fig.\ref{allclte3b} and \ref{allclee1b}.

\begin{figure}[tbh]
\includegraphics[angle=-90,scale=0.5]{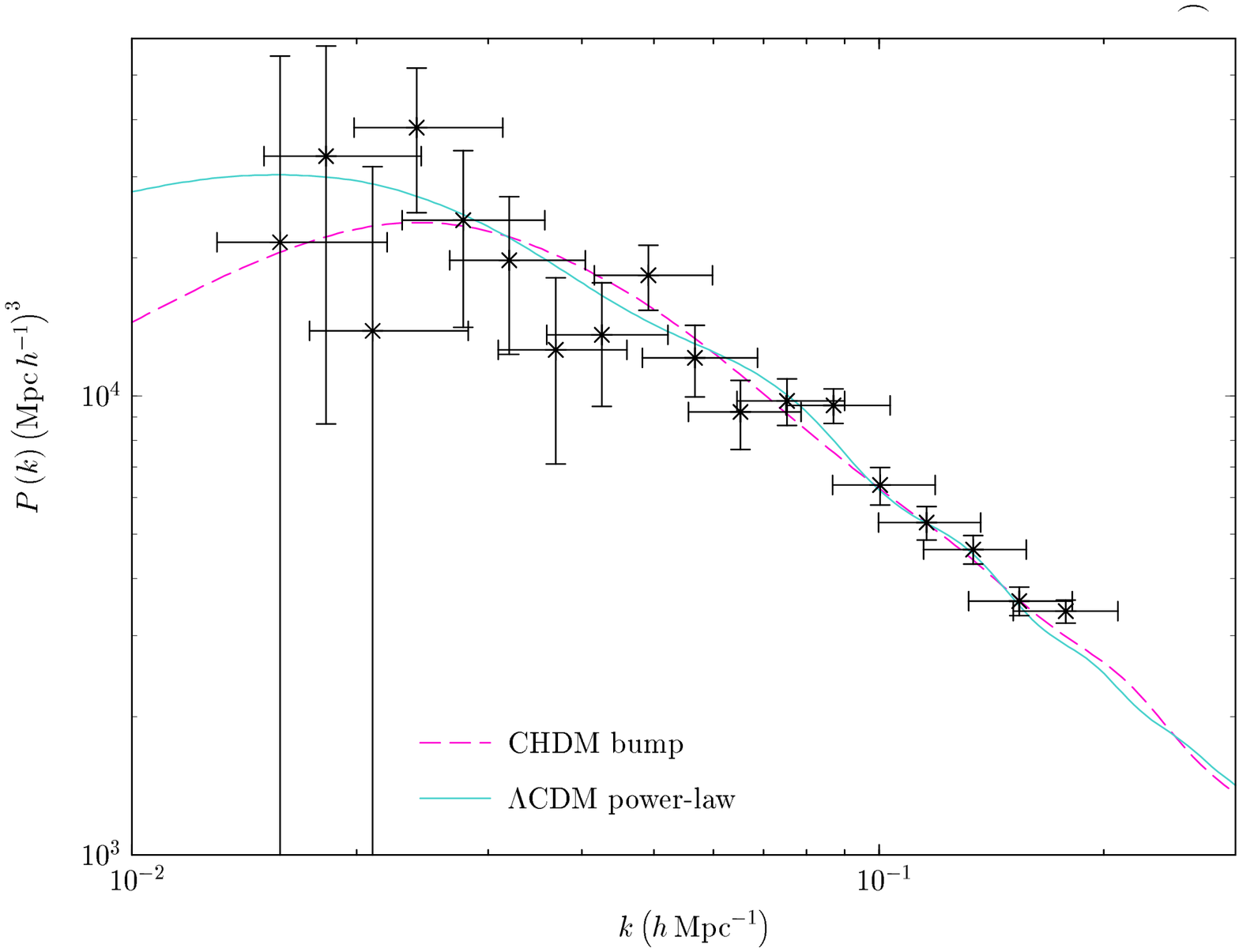}
\caption{\label{allmatpow3b} The best-fit matter power spectra for the
CHDM bump model, with {\em SDSS} data.}
\end{figure}

By adding a hot dark matter component in the form of massive
neutrinos, the amplitude of the matter power spectrum on small scales
is suppressed relative to a pure CDM model and a good fit obtained to
{\sl SDSS} data (see Fig.\ref{allmatpow3b}). As noted earlier
\cite{Blanchard:2003du}, this suppresses $\sigma_8$ so as to provide
better agreement with the value deduced from clusters and weak
lensing. A further suppression occurs due to the downward step in the
primordial spectrum.

\begin{figure}[tbh]
\includegraphics[angle=0,scale=0.85]{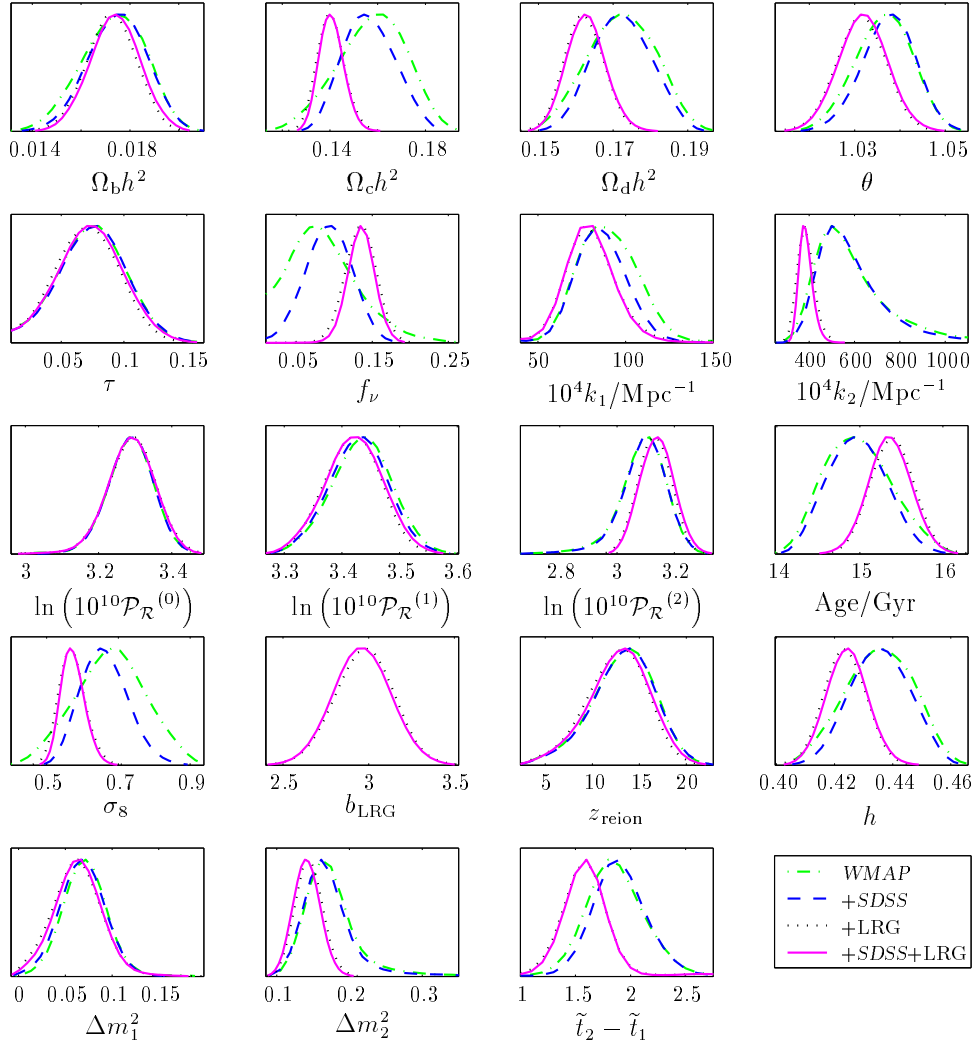}
\caption{\label{all2Edegen4}
 1-dimensional likelihood distributions of the marginalised
 cosmological parameters for the CHDM bump model using the
 {\sl WMAP}, {\sl SDSS} and LRG data.}
\end{figure}

\begin{figure}[tbh]
\includegraphics[angle=-90,scale=0.5]{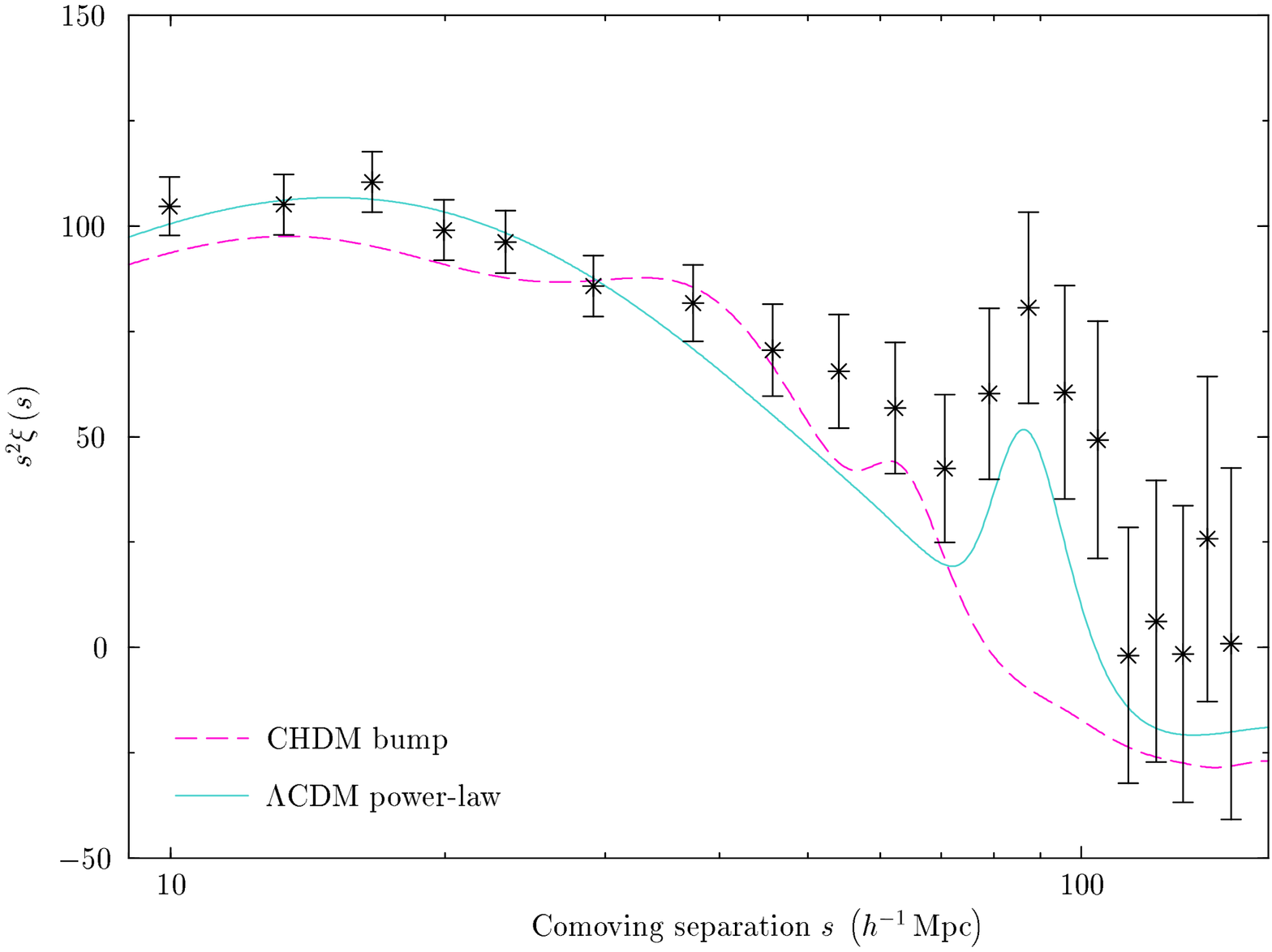}
\caption{\label{allcorr3c} The best-fit two-point galaxy correlation
  functions in the CHDM bump model, with LRG data. The best-fit
  $\Lambda$CDM power-law model, appropriately rescaled, is shown for
  comparison.}
\end{figure}

The constraints on the marginalised cosmological parameters are given
in Table~\ref{edes1} with the order of the non-renormalisable terms
set to $n_1 = 12$ and $n_2 =13$ (which we nevertheless count as 2
additional parameters). We also allow $n_1$ and $n_2$ to vary
continuously (but find no further improvement), with the results shown
in Table~\ref{edes2} and as 1-D likelihood distributions in
Fig.\ref{all2Edegen4}. The fit of this model to the {\sl WMAP} data
and the {\sl SDSS} matter power spectrum is just as good as that of
the power-law $\Lambda$CDM model, as indicated by the $\chi^2$ values
and the {\em vanishing} $\Delta_\mathrm{AIC}$. However, as has been
noted already \cite{Blanchard:2005ev}, the CHDM model does not fit the
LRG two-point correlation function well. This is because the BAO peak
corresponds to the comoving sound horizon at baryon decoupling; the
latter is a function of $\Omega_\mathrm{m}$ and is too low in an
Einstein-de Sitter universe as seen in Fig.\ref{allcorr3c}.

\begin{table}
\begin{center}
\begin{tabular}{|c|c|c|c|c|}
\hline 
&
{\sl WMAP}&
+{\sl SDSS}&
+LRG&
+{\sl SDSS}+LRG\\
\hline
\hline 
$\Omega_\mathrm{b}h^2$&
$0.01748_{-0.00071}^{+0.00073}$&
$0.01762_{-0.00078}^{+0.00080}$&
$0.01692_{-0.00047}^{+0.00047}$&
$0.01688_{-0.00045}^{+0.00044}$\\
\hline 
$\theta$&
$1.0365_{-0.0051}^{+0.0051}$&
$1.0378_{-0.0049}^{+0.0049}$&
$1.0300_{-0.0040}^{+0.0040}$&
$1.0300_{-0.0039}^{+0.0039}$\\
\hline 
$\tau$&
$0.078_{-0.011}^{+0.012}$&
$0.079_{-0.012}^{+0.012}$&
$0.071_{-0.011}^{+0.011}$&
$0.071_{-0.011}^{+0.012}$\\
\hline 
$f_\nu$&
$0.096_{-0.023}^{+0.017}$&
$0.103_{-0.011}^{+0.011}$&
$0.1360_{-0.0092}^{+0.0092}$&
$0.1353_{-0.0067}^{+0.0075}$\\
\hline 
$10^{4}k_1/\mathrm{Mpc}^{-1}$&
$86_{-13}^{+15}$&
$82_{-9.8}^{+11}$&
$77_{-10}^{+12}$&
$77_{-9.5}^{+11}$\\
\hline
$10^{4}k_2/\mathrm{Mpc}^{-1}$&
$527_{-78}^{+78}$&
$539_{-82}^{+84}$&
$380_{-24}^{+24}$&
$379_{-22}^{+22}$\\
\hline
$\ln\left(10^{10}\mathcal{P_{R}}^{(0)}\right)$&
$3.282_{-0.047}^{+0.047}$&
$3.276_{-0.046}^{+0.045}$&
$3.270_{-0.046}^{+0.046}$&
$3.270_{-0.047}^{+0.046}$\\
\hline
$b_\mathrm{LRG}$&
&
&
$2.99_{-0.16}^{+0.16}$&
$2.99_{-0.16}^{+0.16}$\\
\hline
\hline 
$\Omega_\mathrm{c}h^2$&
$0.155_{-0.011}^{+0.012}$&
$0.1539_{-0.0083}^{+0.0084}$&
$0.1387_{-0.0044}^{+0.0041}$&
$0.1387_{-0.0036}^{+0.0037}$\\
\hline 
$\Omega_\mathrm{d}h^2$&
$0.1712_{-0.0062}^{+0.0063}$&
$0.1715_{-0.0059}^{+0.0061}$&
$0.1605_{-0.0031}^{+0.0030}$&
$0.1604_{-0.0030}^{+0.0030}$\\
\hline
$\mathrm{Age/GYr}$&
$15.01_{-0.27}^{+0.27}$&
$14.99_{-0.27}^{+0.26}$&
$15.48_{-0.14}^{+0.14}$&
$15.48_{-0.14}^{+0.14}$\\
\hline
$\sigma_8$&
$0.668_{-0.089}^{+0.093}$&
$0.648_{-0.054}^{+0.053}$&
$0.565_{-0.033}^{+0.032}$&
$0.565_{-0.028}^{+0.029}$\\
\hline
$z_\mathrm{reion}$&
$13.6_{-3.1}^{+3.1}$&
$13.6_{-3.1}^{+3.0}$&
$12.7_{-3.1}^{+3.1}$&
$12.7_{-3.2}^{+3.1}$\\
\hline
$h$&
$0.4344_{-0.0077}^{+0.0078}$&
$0.4348_{-0.0076}^{+0.0079}$&
$0.4212_{-0.0038}^{+0.0037}$&
$0.4211_{-0.0038}^{+0.0038}$\\
\hline
$\Delta m_1^2$&
$0.07476_{-0.00071}^{+0.00070}$&
$0.07468_{-0.00069}^{+0.00068}$&
$0.07459_{-0.00069}^{+0.00068}$&
$0.07459_{-0.00070}^{+0.00068}$\\
\hline
$\Delta m_2^2$&
$0.1510_{-0.0013}^{+0.0013}$&
$0.1508_{-0.0013}^{+0.0012}$&
$0.1507_{-0.0013}^{+0.0012}$&
$0.1507_{-0.0013}^{+0.0012}$\\
\hline
$\widetilde{t}_2-\widetilde{t}_1$&
$1.82_{-0.18}^{+0.16}$&
$1.89_{-0.19}^{+0.15}$&
$1.62_{-0.17}^{+0.12}$&
$1.62_{-0.17}^{+0.11}$\\
\hline
\hline
$\chi^2$&
$11247$&
$11265$&
$11297$&
$11315$\\
\hline
$\Delta_\mathrm{AIC}$&
$0$&
$0$&
$28$&
$29$\\
\hline
\end{tabular}
\end{center}
\caption{\label{edes1} 
 $1 \sigma$ constraints on the marginalised cosmological parameters
 for the CHDM bump model, with the changes in the inflaton mass set by
 choosing $n_1=12$, $n_2=13$ (which are still counted as 2 additional
 free parameters). The 8 parameters in the upper part of the Table are
 varied by CosmoMC, while those in the lower part are derived
 quantities. The $\chi^2$ of the fit is given, as is the Akaike
 information criterion relative to the power-law $\Lambda$CDM model in
 Table~\ref{tab:tilt}.}
\end{table}

\begin{table}
\begin{center}
\begin{tabular}{|c|c|c|c|c|}
\hline 
&
{\sl WMAP}&
+{\sl SDSS}&
+LRG&
+{\sl SDSS}+LRG\\
\hline
\hline 
$\Omega_{\mathrm b}h^{2}$&
$0.0173_{-0.0013}^{+0.0013}$&
$0.0175_{-0.0012}^{+0.0012}$&
$0.01724_{-0.00097}^{+0.00098}$&
$0.01739_{-0.00099}^{+0.00098}$\\
\hline 
$\theta$&
$1.0364_{-0.0062}^{+0.0061}$&
$1.0374_{-0.0056}^{+0.0056}$&
$1.0314_{-0.0052}^{+0.0053}$&
$1.0319_{-0.0052}^{+0.0052}$\\
\hline 
$\tau$&
$0.075_{-0.012}^{+0.012}$&
$0.075_{-0.012}^{+0.012}$&
$0.071_{-0.012}^{+0.012}$&
$0.073_{-0.012}^{+0.012}$\\
\hline 
$f_\nu$&
$0.088_{-0.025}^{+0.016}$&
$0.094_{-0.014}^{+0.015}$&
$0.1349_{-0.0085}^{+0.0088}$&
$0.1354_{-0.0070}^{+0.0073}$\\
\hline 
$10^{4}k_{1}/\mathrm{Mpc}^{-1}$&
$88_{-15}^{+17}$&
$85_{-13}^{+14}$&
$83_{-16}^{+11}$&
$80_{-13}^{+12}$\\
\hline
$10^{4}k_{2}/\mathrm{Mpc}^{-1}$&
$586_{-90}^{+36}$&
$585_{-82}^{+36}$&
$385_{-27}^{+28}$&
$386_{-27}^{+27}$\\
\hline
$\ln\left(10^{10}\mathcal{P}_{\mathcal{R}}^{\left(0\right)}\right)$&
$3.285_{-0.057}^{+0.059}$&
$3.287_{-0.057}^{+0.058}$&
$3.289_{-0.060}^{+0.061}$&
$3.290_{-0.059}^{+0.061}$\\
\hline
$\ln\left(10^{10}\mathcal{P}_{\mathcal{R}}^{\left(1\right)}\right)$&
$3.435_{-0.050}^{+0.050}$&
$3.429_{-0.049}^{+0.048}$&
$3.418_{-0.048}^{+0.047}$&
$3.421_{-0.049}^{+0.049}$\\
\hline
$\ln\left(10^{10}\mathcal{P}_{\mathcal{R}}^{\left(2\right)}\right)$&
$3.088_{-0.067}^{+0.073}$&
$3.091_{-0.067}^{+0.071}$&
$3.132_{-0.056}^{+0.056}$&
$3.140_{-0.058}^{+0.057}$\\
\hline
$b_\mathrm{LRG}$&
&
&
$2.97_{-0.16}^{+0.16}$&
$2.96_{-0.16}^{+0.16}$\\
\hline
\hline 
$\Omega_{\mathrm c}h^{2}$&
$0.157_{-0.014}^{+0.014}$&
$0.156_{-0.011}^{+0.011}$&
$0.1404_{-0.0055}^{+0.0054}$&
$0.1407_{-0.0050}^{+0.0049}$\\
\hline 
$\Omega_{\mathrm d}h^{2}$&
$0.1723_{-0.0090}^{+0.0090}$&
$0.1724_{-0.0077}^{+0.0079}$&
$0.1622_{-0.0050}^{+0.0051}$&
$0.1628_{-0.0050}^{+0.0050}$\\
\hline
$\mathrm{Age/GYr}$&
$14.98_{-0.40}^{+0.40}$&
$14.97_{-0.35}^{+0.35}$&
$15.39_{-0.25}^{+0.25}$&
$15.36_{-0.25}^{+0.25}$\\
\hline
$\sigma_{8}$&
$0.678_{-0.099}^{+0.097}$&
$0.662_{-0.064}^{+0.063}$&
$0.572_{-0.036}^{+0.035}$&
$0.574_{-0.031}^{+0.031}$\\
\hline
$z_{\mathrm{reion}}$&
$13.2_{-3.3}^{+3.2}$&
$13.2_{-3.2}^{+3.2}$&
$12.5_{-3.1}^{+3.1}$&
$12.8_{-3.2}^{+3.1}$\\
\hline
$h$&
$0.435_{-0.012}^{+0.012}$&
$0.436_{-0.010}^{+0.010}$&
$0.4236_{-0.0068}^{+0.0070}$&
$0.4244_{-0.0068}^{+0.0068}$\\
\hline
$\Delta m_{1}$&
$0.072_{-0.022}^{+0.021}$&
$0.068_{-0.023}^{+0.023}$&
$0.062_{-0.025}^{+0.024}$&
$0.064_{-0.024}^{+0.023}$\\
\hline
$\Delta m_{2}$&
$0.176_{-0.030}^{+0.025}$&
$0.172_{-0.028}^{+0.023}$&
$0.145_{-0.018}^{+0.018}$&
$0.141_{-0.018}^{+0.018}$\\
\hline
$\widetilde{t}_2-\widetilde{t}_1$&
$1.91_{-0.28}^{+0.23}$&
$1.93_{-0.24}^{+0.21}$&
$1.61_{-0.21}^{+0.15}$&
$1.63_{-0.21}^{+0.13}$\\
\hline
\hline
$\chi^{2}$&
$11247$&
$11265$&
$11296$&
$11315$\\
\hline
$\Delta_\mathrm{AIC}$&
$0$&
$0$&
$27$&
$29$\\
\hline
\end{tabular}
\end{center}
\caption{\label{edes2} 
 $1 \sigma$ constraints on the marginalised cosmological parameters
 for the CHDM bump model, with the changes in the inflaton mass
 allowed to vary freely. The 10 parameters in the upper part of the
 Table are varied by CosmoMC, while those in the lower part are
 derived quantities. The $\chi^2$ of the fit is given, as is the
 Akaike information criterion relative to the power-law $\Lambda$CDM
 model in Table~\ref{tab:tilt}.}
\end{table}

\section{Parameter degeneracies}

If two different parameters have similar effects on an observable then
they will be negatively correlated in a $2$-D likelihood plot since
the effect of increasing the value of one of them can be undone by
decreasing the value of the other. On the other hand, if two
parameters have opposite effects then increasing the value of one can
compensate for increasing the value of the other and the parameters
will be positively correlated. Such correlations are known as
parameter degeneracies and they limit the constraints which the data
can place on the parameter values.

\subsection{$\Lambda$CDM `step' model degeneracies}
 
The results indicate a strong parameter degeneracy amongst the models
with one phase transition. Moving in parameter space along the
direction of the degeneracy, $\Delta m^2$ increases, while
$\mathcal{P_R}^{(0)}$ and $k_1$ both decrease. This arises because the
amplitude of the TT spectrum on the scale of the acoustic peaks is
governed by the combination $\mathcal{P_R}^{(1)}\mathrm{e}^{-2\tau}$,
where $\mathcal{P_R}^{(1)}$ is the amplitude of the primordial
spectrum after the phase transition:
\begin{equation}
P_{\mathcal{R}}^{(1)} = 
\frac{P_{\mathcal{R}}^{(0)}}{\left(1 - \Delta m^2\right)^2}.
\label{degen1}
\end{equation}
The fit to {\sl WMAP} data thus requires $\mathcal{P_R}^{(1)} \propto
\mathrm{e}^{2\tau}$. Consequently $\tau$ at fixed $\Delta m^2$ is
positively correlated with $\mathcal{P_R}^{(0)}$ (reflecting the
well-known degeneracy between the optical depth and the normalisation
of the TT spectrum). However $\tau$ falls going from $n=15$ to $n=17$
because decreasing $\tau$ reduces $P_\mathcal{R}^{(0)}$, while
increasing $\Delta m^2$ raises $P_\mathcal{R}^{(1)}$.

\begin{figure}[tbh]
\includegraphics[angle=-90,scale=0.5]{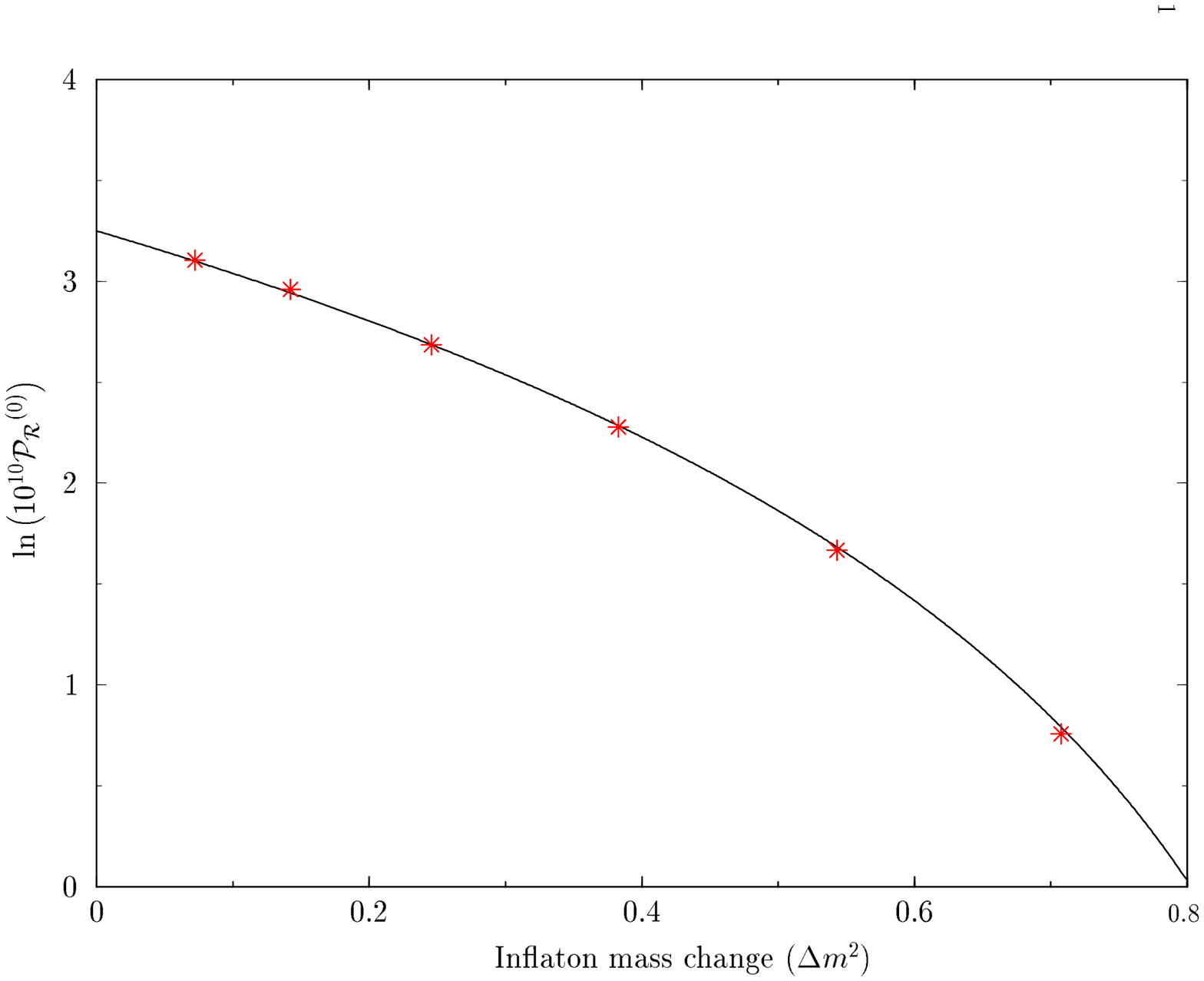}
\caption{\label{all1degen1} 
The degeneracy between the large-scale amplitude of the primordial
 spectrum and the fractional change in the inflaton mass according to
 eq.(\ref{degen1}), with the `data' points taken from
 Table~\ref{tab:all1step}.}
\end{figure}

The value of $\mathcal{P_R}^{(1)}$ is constrained after marginalising
over $\tau$, which means that the relationship between $\Delta m^2$
and $\mathcal{P_R}^{(0)}$ is fitted well by
\begin{equation}
P_{\mathcal{R}}^{(0)} = P_{\mathcal{R}}^{\rm HZ}\left(1 - \Delta m^2\right)^2,
\label{prm}
\end{equation}
where $P_{\mathcal{R}}^{\rm HZ}$ is a constant (being the amplitude of
a scale-invariant Harrison-Zeldovich primordial spectrum). This is
illustrated in Fig.\ref{all1degen1} using results from
Table~\ref{tab:all1step}.

The parameter $k_1$ is negatively correlated with $\Delta m^2$ because
the higher cosmic variance at low $\ell$ allows the data to
accommodate more prominent features in the primordial spectrum at
smaller $k$. The likelihood distribution of $k_1$ is strongly
non-Gaussian, as shown in Fig.\ref{all1Lmany2a}. Each distribution has
maxima wherever the oscillations in the corresponding TT spectrum due
to the phase transition match the glitches in the {\sl WMAP}
measurements.

\subsection{CHDM `bump' model degeneracies}  

\begin{figure}[tbh]
\includegraphics[angle=0,scale=0.8]{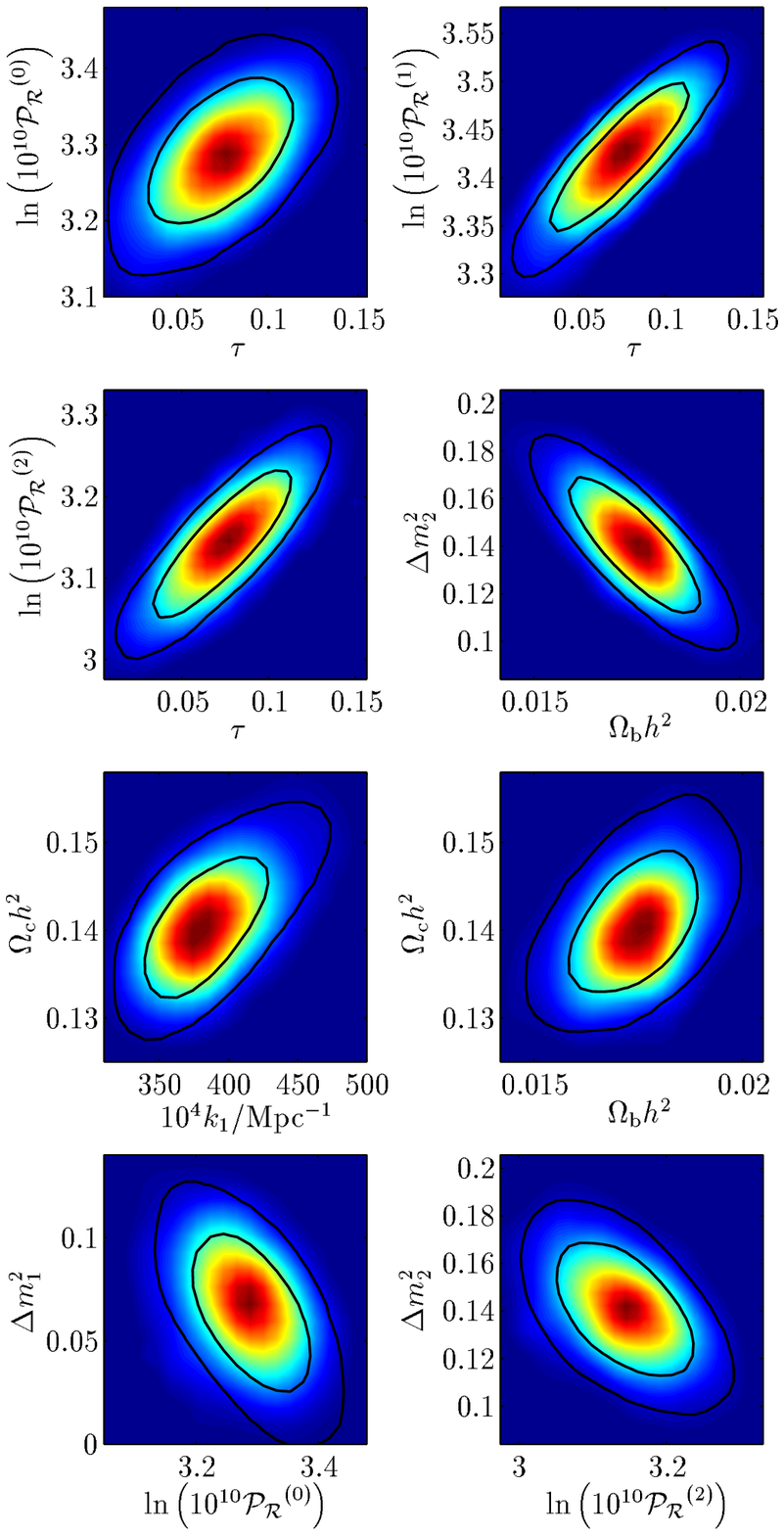}
\caption{\label{all2degen3} 
 Degeneracies between the marginalised cosmological parameters for  
 the CHDM bump model.}
\end{figure}

The CHDM bump model has several parameter degeneracies which are
illustrated in Fig.\ref{all2degen3}. There is a positive correlation
between the optical depth and the amplitude of the primordial spectrum
on medium and small scales.  The correlation between $\tau$ and
$P_{\mathcal{R}}^{(1)}$ is weaker because reionisation does not damp
the TT spectrum on large scales.

Increasing the baryon density increases the height of the first
acoustic peak relative to the second one \cite{Hu:1995en}, while
increasing $\Delta m^2_2$ increases the size of the second step in the
primordial spectrum and so boosts the height of the first peak
relative to the other peaks. Thus $\omega_{\rm b}$ and $\Delta m_2^2$
are negatively correlated since both parameters have the same effect
on the height of the first peak relative to the second \cite{Hu:1995en}.

The bump in the primordial spectrum becomes broader when $k_1$ is
increased, which increases the height of the second acoustic peak
relative to the third. Reducing the density of CDM has a similar
effect on the peak heights, hence $k_1$ and $\omega_{\rm c}$ are
positively correlated. The baryon and CDM densities are also
positively correlated because increasing $\omega_{\rm b}$ increases
the height of the first acoustic peak, while increasing $\omega_{\rm
c}$ has the opposite effect \cite{Hu:1995en}.

The error bars of the {\sl WMAP} data are smallest in the region of
the first acoustic peak, so $P_{\mathcal{R}}^{(1)}$ is more tightly
constrained than $P_{\mathcal{R}}^{(0)}$ or $P_{\mathcal{R}}^{(2)}$,
as is apparent from Table~\ref{edes2}. Consequently the parameters
$P_{\mathcal{R}}^{(0)}$ and $\Delta m_1^2$ satisfy
\begin{equation}
P_{\mathcal{R}}^{(0)} = P_{\mathcal{R}}^{\mathrm{HZ}}
\left(1 - \Delta m_1^2\right)^2, 
\end{equation}
which is similar to eq.(\ref{prm}).

From the last equality of eq.(\ref{ampl}) it is apparent that increasing
$\Delta m_2^2$ reduces $\mathcal{P_R}^{(2)}$, hence these parameters
are anti-correlated, as seen in the final panel of
Fig.\ref{all2degen3}.

\section{Conclusions}

Anisotropies in the CMB and correlations in the large-scale
distribution of galaxies reflect the primordial perturbations,
presumably from inflation, but convoluted with the {\em a priori}
unknown effects of their evolution in matter. Thus the values of the
parameters of the assumed cosmological model are necessarily uncertain
due to our limited knowledge of both the dark matter content of the
universe, as well as of the nature of the primordial
perturbations. Nevertheless, by combining various lines of evidence,
it has been possible to make a strong case for the concordance
$\Lambda$CDM model, based on a FRW cosmology
\cite{Peebles:2002gy}. This model appears deceptively simple, but
invokes vacuum energy at an unnatural scale, $\rho_\Lambda \sim
(10^{-30}M_\mathrm{P})^4$, as the dominant constituent of the
universe. There is no satisfactory understanding of this from
fundamental theoretical considerations
\cite{Weinberg:1988cp,Nobbenhuis:2004wn}. It is therefore worth
reexamining whether the precision {\sl WMAP} data, which has played a
key role in the general acceptance of the $\Lambda$CDM model, cannot
be otherwise interpreted.

In this work we have focussed on the fact that our present
understanding of the physics of inflation is rather primitive and
there is no compelling reason why the generated primordial adiabatic
scalar density perturbation should be close to the scale-invariant
Harrison-Zeldovich form, as is usually assumed. This is indeed what is
expected in the simplest toy model of an inflaton field evolving
slowly down a suitably flat potential, but attempts to realise this in
a physical theory such as supergravity or string/M-theory are plagued
with difficulties. Such considerations strongly suggest moreover that
even if just one scalar field comes to dominate the energy density and
drives inflation, the non-trivial dynamics of other scalar fields in
the inflating universe can affect its slow-roll evolution and thus
create features in the primordial spectrum. Indeed there is some
indication for such spectral features in the {\sl WMAP}
observations. Whether they are real rather than systematic effects
will be tested further in the {\sl WMAP} 5-yr analysis and by the
forthcoming {\sl Planck} mission \cite{unknown:2006uk}. It has also
been noted that significant non-Gaussianity would be generated when
there are sharp features in the inflaton potential
\cite{Chen:2006xj}. This can be computed in our model and would
constitute an independent test.

We have demonstrated that such spectral features can have a major
impact on cosmological parameter estimation, in that the {\sl WMAP}
data can be fitted without requiring any dark energy, if there happens
to be a mild bump in the primordial spectrum on spatial scales of
order the Hubble radius at the last scattering surface. This has been
modelled here using the supergravity-based multiple inflation model
\cite{Adams:1997de} with reasonable choices for the model
parameters. Although other possibilities for generating spectral
features have been investigated, our framework has the advantage of
being based on effective field theory and thus calculable. 

We have also departed from earlier work on parameter estimation in
considering possible values of the Hubble constant well below the HKP
value \cite{Freedman:2000cf} which is usually input as a prior.
Moreover the CHDM `bump' model cannot fit the luminosity distance
measurements of SN~Ia, since the supernovae are fainter than predicted
in a {\em homogeneous} Einstein-de Sitter cosmology.  However, it has
been noted that deep measurements using physical methods yield a lower
value of $h$ than that measured locally
\cite{Freedman:2000cf,Blanchard:2003du} as would be the case if we are
located in an underdense void which is expanding faster than the
average \cite{Zehavi:1998gz}. There are ongoing attempts to test
whether we are indeed inside such a `Hubble bubble'
\cite{Jha:2006fm,Conley:2007ng}. This is enormously important since
such an inhomogeneous LTB cosmology can in principle address the
failures of our FRW framework, with regard to the SN~Ia Hubble diagram
and the BAO peak \cite{Biswas:2006ub}.

Moreover we have shown that the observed power spectrum of galaxy
clustering can also be accounted for by modifying the nature of the
dark matter, in particular allowing for a significant component due to
massive neutrinos. Whether neutrinos do have the required mass of
$\sim 0.5$~eV will be definitively tested in the forthcoming {\sl
KATRIN} experiment \cite{Drexlin:2004as}. The position of the baryon
peak seen in the galaxy correlation function may also be matched in
the LTB framework, with a low global Hubble constant
\cite{Biswas:2006ub}.

Regardless of whether this model is the right description of the
physical universe, we wish to emphasise that it provides a {\em good}
fit to the {\sl WMAP} data. Thus for progress to be made in pinning
down the cosmological parameters and extending our understanding of
inflation using precision CMB data, it is clearly necessary for a
broader analysis framework to be adopted than has been the practice so
far.

\begin{acknowledgments}
  We thank Graham Ross for discussions concerning inflation and Jo
  Dunkley and Alex Lewis for advice on MCMC codes. We are grateful to
  the {\sl WMAP} team for making their data and analysis tools
  publicly available, and to the Referee for helpful suggestions. This
  work was supported by a STFC Senior Fellowship award (PPA/C506205/1)
  and by the EU Marie Curie Network ``UniverseNet''
  (HPRN-CT-2006-035863).
\end{acknowledgments}

\appendix

\section{\label{sampling}adaptive sampling algorithm}

\begin{figure}[tbh]
\includegraphics[angle=-90,scale=0.40]{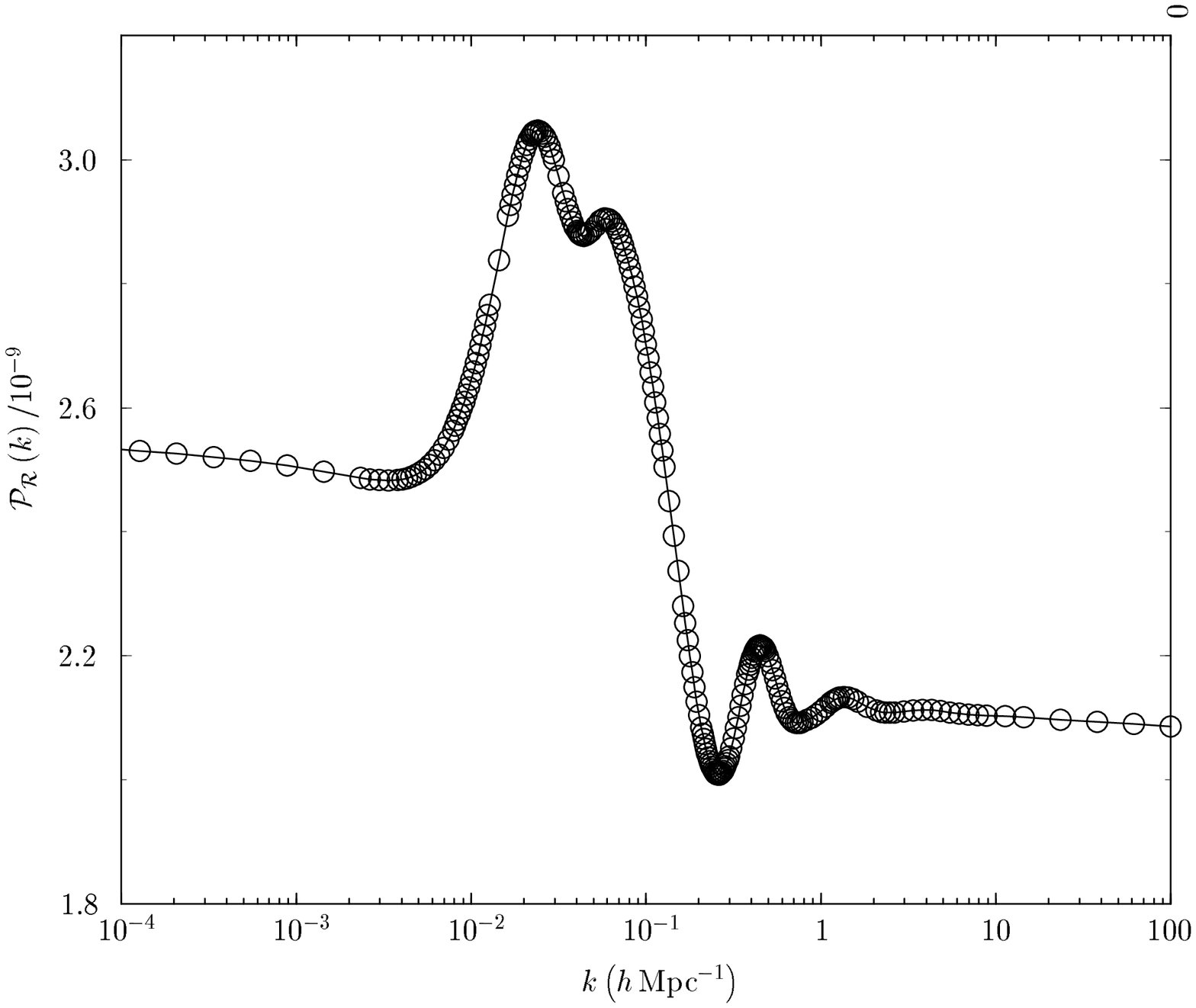}
\caption{\label{adapt} 
 An example result from our adaptive sampling algorithm which samples
 the power spectrum more finely where the curvature is higher.}
\end{figure}

For plotting or interpolating a function $y = f(x)$ over an interval
$(x_\mathrm{min}, x_\mathrm{max})$ it is useful to have an algorithm
that generates a set of points $\left\{x_i, f\left(x_i\right)\right\}$, 
where $x_{i+1} > x_i$, such that the
density of the samples $x_{i}$ in the abscissa increases with
$\mathrm{d}^2 f/\mathrm{d}x^2$. One such algorithm is listed here. It
works by ensuring that the value of the function at the mid-point of
each interval $(x_i, x_{i+2})$ is within some tolerance
$\varepsilon$ of the value there linearly interpolated from the
end-points of the interval.
\begin{enumerate}
\item Form two lists $\{x_1,x_2,\ldots,x_N\}$ and 
$\{ y_1,y_2,\ldots,y_N\}$ where the $x_i's$ are equally spaced,
\begin{equation}
x_i=x_{\mathrm{min}}+\left(\frac{i-1}{N-1}\right)x_{\mathrm{max}},
\quad i=1,2,\ldots,N,
\end{equation}
and $y_i=f\left(x_i\right)$.
\item Start with the first interval $(x_1,x_2).$
\item For the interval $(x_i,x_{i+1})$ under
consideration calculate $x_m \equiv \left(x_i+x_{i+1}\right)/2$ and
$y_m \equiv f(x_m)$.  Insert $x_m$ between $x_i$ and $x_{i+1}$ in
the list of $x$ values and $y_m$ between $y_i$ and $y_{i+1}$ in the
list of $y$ values.
\item If $\left|y_m-\left(y_i+y_{i+1}\right)/2\right|>\varepsilon$
repeat step $3$ with the new interval $(x_i,x_m)$.
\item Otherwise repeat step $3$ with the next interval
$\left(x_{i+1},x_{i+2}\right)$, unless $x_{i+1}=x_{\rm max}$, in which
case finish.
\end{enumerate}
This algorithm requires a minimum number of function evaluations and
none are wasted. In our work we set $x=\ln\widetilde{k}$ and achieve
satisfactory results with $N=20$ and $\varepsilon=10^{-12}$, as
illustrated in Fig.\ref{adapt}.

\section{\label{mcmc}MCMC likelihood analysis}

We use the Monte Carlo Markov Chain (MCMC) approach to cosmological
parameter estimation. It is a method for drawing samples from the
posterior distribution
$P\left(\mbox{\boldmath$\varpi$}|\mathrm{data}\right)$ of the
parameters $\mbox{\boldmath$\varpi$}$, given the data. According to
Bayes' theorem, the posterior distribution is given by:
\begin{equation}
P\left(\mbox{\boldmath$\varpi$}|\mathrm{data}\right)=\frac{\mathcal{L}
\left(\mathrm{data}|\mbox{\boldmath$\varpi$}\right)P\left(\mbox{\boldmath$\varpi$}
\right)}{\int\mathcal{L}\left(\mathrm{data}|\mbox{\boldmath$\varpi$}\right)P
\left(\mbox{\boldmath$\varpi$}\right)\mathrm{d}\mbox{\boldmath$\varpi$}},
\end{equation} 
where $\mathcal{L}\left(\mathrm{data}|\mbox{\boldmath$\varpi$}\right)$
is the likelihood of the parameters and
$P\left(\mbox{\boldmath$\varpi$}\right)$ is their prior
distribution. We use flat priors on all the parameters listed in
Table~\ref{tab:priors}. With the advent of
massively parallel supercomputers which can run several chains
simultaneously MCMC has become the standard tool for model fitting. At
best, the number of likelihood evaluations required increases linearly
with the number of parameters, $n$.  This is much slower than
traditional grid based methods which need $m^n$ evaluations where $m$
is the number of grid points in each parameter.

We use a modified version of the CosmoMC software package
\cite{Lewis:2002ah}.\footnote{http://cosmologist.info/cosmomc/}
CosmoMC contains an implementation of the Metropolis-Hastings
algorithm for generating Markov chains, which runs as follows: let the
vector $\mbox{\boldmath$\varpi$}^{\left(i\right)}$ be the current
location of the chain in parameter space. A candidate
$\mbox{\boldmath$\widetilde{\varpi}$}$ for the next step in the chain
is chosen from some proposal distribution
$q\left(\mbox{\boldmath$\widetilde{\varpi}$}
\Big|\mbox{\boldmath$\varpi$}^{\left(i\right)}\right)$. The proposed
location $\mbox{\boldmath$\widetilde{\varpi}$}$ may differ from
$\mbox{\boldmath$\varpi$}^{\left(i\right)}$ by either some or all of
the parameter values.  Then
$\mbox{\boldmath$\varpi$}^{\left(i+1\right)}=\mbox{\boldmath$\widetilde{\varpi}$}$
with probability
\begin{equation}
\alpha\left(\mbox{\boldmath$\widetilde{\varpi}$}\Big|\mbox{\boldmath$\varpi$}^{\left(i\right)}
\right)=\min\left[1,\frac{q\left(\mbox{\boldmath$\varpi$}^{\left(i\right)}\Big|\mbox
{\boldmath$\widetilde{\varpi}$}\right)}{q\left(\mbox{\boldmath$\widetilde{\varpi}$}
\Big|\mbox{\boldmath$\varpi$}^{\left(i\right)}\right)}\frac{\mathcal{L}
\left(\mathrm{data}\Big|\mbox{\boldmath$\widetilde{\varpi}$}\right)}{\mathcal{L}
\left(\mathrm{data}|\mbox{\boldmath$\varpi$}^{\left(i\right)}\right)}\right],
\end{equation}
and the step is said to be `accepted', otherwise
$\mbox{\boldmath$\varpi$}^{\left(i+1\right)}=\mbox{\boldmath$\varpi$}^{\left(i\right)}$
and the step is `rejected'. A chain that is built up from many such
steps will be Markovian as the proposal distribution
$q\left(\mbox{\boldmath$\widetilde{\varpi}$}\Big|
\mbox{\boldmath$\varpi$}^{\left(i\right)}\right)$ is a function of
$\mbox{\boldmath$\varpi$}^{\left(i\right)}$ only and not earlier
steps. When
$q\left(\mbox{\boldmath$\widetilde{\varpi}$}\Big|\mbox{\boldmath$\varpi$}^{\left(i\right)}
\right)=q\left(\mbox{\boldmath$\widetilde{\varpi}$}-\mbox{\boldmath$\varpi$}^{\left(i\right)}
\right)$, as in the case of CosmoMC, the chain executes a random walk
in parameter space, apart from the rejected steps. Since $\alpha$
satisfies the reversibility condition,
\begin{equation}
\mathcal{L}\left(\mbox{\boldmath$\varpi$}^{\left(i\right)}\right)q\left(\mbox{\boldmath$
\varpi$}^{\left(i\right)}\Big|\mbox{\boldmath$\widetilde{\varpi}$}\right)\alpha\left(\mbox{
\boldmath$\varpi$}^{\left(i\right)}\Big|\mbox{\boldmath$\widetilde{\varpi}$}\right)=\mathcal
{L}\left(\mbox{\boldmath$\widetilde{\varpi}$}\right)q\left(\mbox{\boldmath$
\widetilde{\varpi}$}\Big|\mbox{\boldmath$\varpi$}^{\left(i\right)}\right)\alpha\left(\mbox
{\boldmath$\widetilde{\varpi}$}\Big|\mbox{\boldmath$\varpi$}^{\left(i\right)}\right),
\end{equation} 
It can be shown by construction that
$\mathcal{L}\left(\mbox{\boldmath$\varpi$} \right)$ will be the
equilibrium distribution of the chain in nearly all practical
situations. Thus when equilibrium is reached the chain represents a
random walk where proportionally more time is spent in regions of
higher likelihood. This occurs after a transient period known as the
`burn in' where the steps are highly correlated with the starting
point (and are usually discarded). Once the chain is equilibrated, it
must run long enough to explore all of the relevent parameter space
after which the chain is said to be `mixed'. A chain which moves
rapidly through parameter space so that this happens quickly is said
to have `good mixing'. Shot noise in the statistics inferred from the
chain is also reduced by running longer.

When the statistics accurately reflect the posterior distribution and
the chain can be used reliably for parameter estimation, it is said to
have `converged'. In practice, convergence is often taken to have
occurred when the results derived from a chain are independent of its
starting point. Many methods of diagnosing convergence have been
proposed. CosmoMC uses one recommended in ref.\cite{gelman} which
involves running several chains with widely dispersed starting
points. (For a different technique used in the context of cosmological
parameter estimation see ref.\cite{Dunkley:2004sv}.) A comparison is
made between the within-chain variance of the chains, which is the
mean of the variance of each of the chains, and the between-chain
variance, which is equal to the length of the chains multiplied by the
variance of the means of the chains (in the case where all chains have
the same length). The idea is that they should be the same after
convergence, as all the chains exhibit the same behaviour and the
dependence on the starting position has been lost. Ref.\cite{gelman}
present a quantity $R$ containing the ratio of the within-chain and
the between-chain variances such that at convergence $R \simeq 1$. In
our work we assume convergence to have occurred when $R=1.02$.

The speed with which a chain converges is strongly dependent on the
proposal distribution. If too many large steps are proposed, the
acceptance rate will be low as most steps will be away from the region
of high likelihood, and the chain will be stationary for long periods
of time. If too many small steps are proposed, most will be accepted
but the chain will move slowly through parameter space. In either case
the chains will mix slowly and the steps will be highly correlated. We
use the results of preliminary MCMC runs to fine-tune the width of the
proposal distribution before proceeding with the final run.

Cosmological parameters in CosmoMC are divided into two classes:
`fast' and `slow'. Slow parameters are those upon which the transfer
function is dependent. All others are fast parameters, including those
which govern the primordial power spectrum or correspond to
calibration uncertainties in the data. Steps in which any of the slow
parameters change take a relatively long time to compute, because the
complicated Einstein-Boltzmann equations for the transfer function
must be solved. Once the transfer function is known however, steps in
which only the fast parameters change can be taken quickly as CAMB
uses linear perturbation theory. Although the calculation of the
primordial power spectrum takes longer than usual in our modified
version of CAMB, a fast step is still quicker than a slow one, and we
retain the fast and slow division of parameters. CosmoMC exploits this
split by alternately taking fast and slow steps, which allows a more
rapid exploration of the parameter space than slow steps alone.

In the absence of a covariance matrix for the parameters, CosmoMC
chooses at random a basis in the slow parameter subspace. It then
proposes in turn a step in the direction of each basis vector with
every subsequent slow step.  Once all the basis vectors have been used
in this way, CosmoMC chooses a new random set and repeats the
cycle. The length of the step is determined by the proposal
distribution, which in CosmoMC is based upon a two dimensional radial
Gaussian function mixed with an exponential. Fast steps are taken in a
similar way, by cycling through random basis vectors in the fast
subspace. This is done to reduce the risk of the chain doubling back
on itself.

When a covariance matrix is available, CosmoMC takes slow steps by
cycling through random bases in the $n_{slow}$ dimensional subspace
spanned by the $n_{slow}$ largest eigenvectors of the covariance
matrix. In this way fast parameters that are correlated with slow ones
are also changed during a slow step, in the direction of the
degeneracies, which increases the mobility of the chain. Fast steps
are just made in the fast subspace as before, and together the fast
and slow steps can traverse the whole parameter space.

Given $N$ samples $\mbox{\boldmath$\varpi$}^{\left(i\right)}$ drawn
from $P\left(\mbox{\boldmath$\varpi$}\right|\textrm{data})$ the best
estimate for the distribution is formally
\begin{equation}
P_{N}\left(\mbox{\boldmath$\varpi$}\right|\textrm{data})=\frac{1}{N}
\sum_{i=1}^{N}\delta\left(\mbox{\boldmath$\varpi$}-\mbox{\boldmath$\varpi$}^{\left(i\right)}\right).
\label{mcmc1}
\end{equation}
 However, rather than studying the full function
$P\left(\mbox{\boldmath$\varpi$}\right|\textrm{data})$ it is often
easier to interpret marginalised probability distributions obtained by
integrating over a subset of the cosmological parameters,
\begin{equation}
P\left(\varpi_{1},\varpi_{2},\ldots,\varpi_{m}\right|\textrm{data})\equiv
\int P\left(\varpi_{1},\varpi_{2},\ldots,\varpi_{n}\right|\textrm{data})\prod_{i=m+1}^{n}\mathrm{d}\varpi_{i}.
\label{mcmc2}
\end{equation}
From eq.(\ref{mcmc1}) the probability given the data that a parameter
lies in a particular interval is proportional to the number of samples
that fall into the interval. Thus the marginalised probability
distributions can be found using histograms of the samples. For $1$-D
marginalised probability distributions confidence intervals are
frequently quoted such that a fraction $1-\alpha$ of the samples fall
within the interval, while $\alpha/2$ lie higher and $\alpha/2$ lie
lower.

The expectation value of a parameter given the data is
\begin{equation}
\left\langle \varpi_{a}\right|\left.\mathrm{data}\right\rangle =
\int\varpi_{a}P\left(\varpi_{a}\right|\textrm{data})\mathrm{d}\varpi_{a}.
\end{equation}
Using eqs.(\ref{mcmc1}) and (\ref{mcmc2}) it is approximated by
\begin{equation}
\left\langle \varpi_{a}\right|\left.\mathrm{data}\right\rangle _{N}=
\frac{1}{N}\sum_{i=1}^{N}\varpi_{a}^{\left(i\right)}.
\end{equation}


\begin{thebibliography}{99}

\bibitem{Spergel:2003cb}
D.~N.~Spergel {\it et al.},
Astrophys.\ J.\ Suppl.\ {\bf 148}, 175 (2003).

\bibitem{Spergel:2006hy}
D.~N.~Spergel {\it et al.} [WMAP Collaboration],
arXiv:astro-ph/0603449.

\bibitem{Efstathiou:2003wr}
G.~Efstathiou,
Mon.\ Not.\ Roy.\ Astron.\ Soc.\  {\bf 346}, L26 (2003);
{\em ibid} {\bf 348}, 885 (2004).

\bibitem{Slosar:2004xj}
A.~Slosar and U.~Seljak,
Phys.\ Rev.\  D {\bf 70}, 083002 (2004)

\bibitem{Park:2006dv}
C.~G.~Park, C.~Park and J.~R.~I.~Gott,
Astrophys.\ J.\ {\bf 660}, 959 (2007).

\bibitem{Hinshaw:2006ia}
G.~Hinshaw {\it et al.} [WMAP Collaboration],
arXiv:astro-ph/0603451.

\bibitem{Peiris:2003ff}
H.~V.~Peiris {\it et al.},
Astrophys.\ J.\ Suppl.\  {\bf 148}, 213 (2003).

\bibitem{Adams:1997de}
J.~A.~Adams, G.~G.~Ross and S.~Sarkar,
Nucl.\ Phys.\ B {\bf 503}, 405 (1997).

\bibitem{Adams:2001vc}
J.~Adams, B.~Cresswell and R.~Easther,
Phys.\ Rev.\ D {\bf 64}, 123514 (2001).

\bibitem{Covi:2006ci}
L.~Covi, J.~Hamann, A.~Melchiorri, A.~Slosar and I.~Sorbera,
Phys.\ Rev.\  D {\bf 74}, 083509 (2006).

\bibitem{Hamann:2007pa}
J.~Hamann, L.~Covi, A.~Melchiorri and A.~Slosar,
Phys.\ Rev.\  D {\bf 76}, 023503 (2007).

\bibitem{German:2001tz}
G.~German, G.~G.~Ross and S.~Sarkar,
Nucl.\ Phys.\  B {\bf 608}, 423 (2001).

\bibitem{Hunt:2004vt}
P.~Hunt and S.~Sarkar,
Phys.\ Rev.\  D {\bf 70}, 103518 (2004).

\bibitem{Langlois:2004px}
D.~Langlois and F.~Vernizzi,
JCAP {\bf 0501}, 002 (2005).

\bibitem{Starobinsky:ts}
A.~A.~Starobinsky,
JETP Lett.\  {\bf 55}, 489 (1992).

\bibitem{Martin:2003sg}
J.~Martin and C.~Ringeval,
Phys.\ Rev.\ D {\bf 69}, 083515 (2004);
{\it ibid} {\bf 69}, 127303 (2004);
JCAP {\bf 0501}, 007 (2005).

\bibitem{Danielsson:2006gg}
U.~H.~Danielsson,
arXiv:astro-ph/0606474.

\bibitem{Mathews:2004vu}
G.~J.~Mathews, D.~J.~H.~Chung, K.~Ichiki, T.~Kajino and M.~Orito,
Phys.\ Rev.\  D {\bf 70}, 083505 (2004).

\bibitem{Bridle:2003sa}
S.~L.~Bridle, A.~M.~Lewis, J.~Weller and G.~Efstathiou,
Mon.\ Not.\ Roy.\ Astron.\ Soc.\  {\bf 342}, L72 (2003).

\bibitem{Hannestad:2003zs}
S.~Hannestad,
JCAP {\bf 0404}, 002 (2004).

\bibitem{Bridges:2005br}
M.~Bridges, A.~N.~Lasenby and M.~P.~Hobson,
Mon.\ Not.\ Roy.\ Astron.\ Soc.\  {\bf 369}, 1123 (2006).

\bibitem{Bridges:2006zm}
M.~Bridges, A.~N.~Lasenby and M.~P.~Hobson,
arXiv:astro-ph/0607404.

\bibitem{Mukherjee:2003ag}
P.~Mukherjee and Y.~Wang,
Astrophys.\ J.\  {\bf 599}, 1 (2003).

\bibitem{Sealfon:2005em}
C.~Sealfon, L.~Verde and R.~Jimenez,
Phys.\ Rev.\  D {\bf 72}, 103520 (2005).

\bibitem{Leach:2005av}
S.~Leach,
Mon.\ Not.\ Roy.\ Astron.\ Soc.\  {\bf 372}, 646 (2006).

\bibitem{Shafieloo:2003gf}
A.~Shafieloo and T.~Souradeep,
Phys.\ Rev.\ D {\bf 70}, 043523 (2004).

\bibitem{Shafieloo:2006hs}
A.~Shafieloo, T.~Souradeep, P.~Manimaran, P.~K.~Panigrahi and R.~Rangarajan,
Phys.\ Rev.\  D {\bf 75}, 123502 (2007).

\bibitem{Kogo:2003yb}
N.~Kogo, M.~Matsumiya, M.~Sasaki and J.~Yokoyama,
Astrophys.\ J.\ {\bf 607}, 32 (2004).

\bibitem{Kogo:2004vt}
N.~Kogo, M.~Sasaki and J.~Yokoyama,
Phys.\ Rev.\  D {\bf 70}, 103001 (2004);
Prog.\ Theor.\ Phys.\  {\bf 114}, 555 (2005).

\bibitem{Tocchini-Valentini:2004ht}
D.~Tocchini-Valentini, M.~Douspis and J.~Silk,
Mon.\ Not.\ Roy.\ Astron.\ Soc.\  {\bf 359}, 31 (2005)

\bibitem{Tocchini-Valentini:2005ja}
D.~Tocchini-Valentini, Y.~Hoffman and J.~Silk,
Mon.\ Not.\ Roy.\ Astron.\ Soc.\ {\bf 367}, 1095 (2006).

\bibitem{Freedman:2000cf}
W.~L.~Freedman {\it et al.},
Astrophys.\ J.\  {\bf 553}, 47 (2001).

\bibitem{Tegmark:2003uf}
M.~Tegmark {\it et al.}  [SDSS Collaboration],
Astrophys.\ J.\  {\bf 606}, 702 (2004)

\bibitem{Elgaroy:2003yh}
O.~Elgaroy and O.~Lahav,
JCAP {\bf 0304}, 004 (2003).

\bibitem{Blanchard:2003du}
A.~Blanchard, M.~Douspis, M.~Rowan-Robinson and S.~Sarkar,
Astron.\ Astrophys.\  {\bf 412}, 35 (2003).

\bibitem{Yao:2006px}
W.~M.~Yao {\it et al.}  [Particle Data Group],
J.\ Phys.\ G {\bf 33}, 1 (2006).

\bibitem{Blanchard:2005ev}
A.~Blanchard, M.~Douspis, M.~Rowan-Robinson and S.~Sarkar,
Astron.\ Astrophys.\  {\bf 449}, 925 (2006).

\bibitem{Eisenstein:2005su}
D.~J.~Eisenstein {\it et al.}  [SDSS Collaboration],
Astrophys.\ J.\  {\bf 633}, 560 (2005).

\bibitem{Biswas:2006ub}
T.~Biswas, R.~Mansouri and A.~Notari,
arXiv:astro-ph/0606703.

\bibitem{Goodwin:1999ej}
S.~P.~Goodwin, P.~A.~Thomas, A.~J.~Barber, J.~Gribbin and L.~I.~Onuora,
arXiv:astro-ph/9906187.

\bibitem{Celerier:1999hp}
M.~N.~Celerier,
Astron.\ Astrophys.\  {\bf 353}, 63 (2000).

\bibitem{Tomita:2000jj}
K.~Tomita,
Mon.\ Not.\ Roy.\ Astron.\ Soc.\  {\bf 326}, 287 (2001).

\bibitem{Alnes:2005rw}
H.~Alnes, M.~Amarzguioui and O.~Gron,
Phys.\ Rev.\  D {\bf 73}, 083519 (2006).

\bibitem{Enqvist:2006cg}
K.~Enqvist and T.~Mattsson,
JCAP {\bf 0702}, 019 (2007).

\bibitem{Alnes:2006uk}
H.~Alnes and M.~Amarzguioui,
Phys.\ Rev.\  D {\bf 75}, 023506 (2007).

\bibitem{Celerier:2007jc}
M.~N.~Celerier,
arXiv:astro-ph/0702416.

\bibitem{Randall:1997kx}
L.~Randall,
arXiv:hep-ph/9711471.

\bibitem{Lyth:1998xn}
D.~H.~Lyth and A.~Riotto,
Phys.\ Rept.\  {\bf 314}, 1 (1999)

\bibitem{Mukhanov:1990me}
V.~F.~Mukhanov, H.~A.~Feldman and R.~H.~Brandenberger,
Phys.\ Rept.\ {\bf 215}, 203 (1992).

\bibitem{Kinney:2006qm}
W.~H.~Kinney, E.~W.~Kolb, A.~Melchiorri and A.~Riotto,
Phys.\ Rev.\  D {\bf 74}, 023502 (2006).

\bibitem{Lewis:1999bs}
A.~Lewis, A.~Challinor and A.~Lasenby,
Astrophys.\ J.\  {\bf 538}, 473 (2000).

\bibitem{Eisenstein:1997ik}
D.~J.~Eisenstein and W.~Hu,
Astrophys.\ J.\  {\bf 496}, 605 (1998)

\bibitem{Smith:2002dz}
R.~E.~Smith {\it et al.}  [The Virgo Consortium Collaboration],
Mon.\ Not.\ Roy.\ Astron.\ Soc.\  {\bf 341}, 1311 (2003).

\bibitem{Hamilton:1999uv}
A.~J.~S.~Hamilton,
Mon.\ Not.\ Roy.\ Astron.\ Soc.\  {\bf 312}, 257 (2000).

\bibitem{Kaiser:1987qv}
N.~Kaiser,
Mon.\ Not.\ Roy.\ Astron.\ Soc.\  {\bf 227}, 1 (1987).

\bibitem{Hamilton:1991es}
A.~J.~S.~Hamilton, A.~Matthews, P.~Kumar and E.~Lu,
Astrophys.\ J.\  {\bf 374}, L1 (1991).

\bibitem{Akaike:1974}
H.~Akaike,
IEEE\ Trans.\ Auto.\ Control, {\bf 19}, 716 (1974).	

\bibitem{Liddle:2004nh}
A.~R.~Liddle,
Mon.\ Not.\ Roy.\ Astron.\ Soc.\  {\bf 351}, L49 (2004).

\bibitem{Silk:2000sn}
J.~Silk and E.~Gawiser,
Phys.\ Scripta {\bf T85}, 132 (2000).

\bibitem{Hu:1995en}
W.~Hu and N.~Sugiyama,
Astrophys.\ J.\  {\bf 471}, 542 (1996)

\bibitem{Peebles:2002gy}
P.~J.~E.~Peebles and B.~Ratra,
Rev.\ Mod.\ Phys.\  {\bf 75}, 559 (2003).

\bibitem{Weinberg:1988cp}
S.~Weinberg,
Rev.\ Mod.\ Phys.\  {\bf 61}, 1 (1989);
arXiv:astro-ph/0005265.

\bibitem{Nobbenhuis:2004wn}
S.~Nobbenhuis,
Found.\ Phys.\  {\bf 36}, 613 (2006).

\bibitem{unknown:2006uk}
Planck Collaboration,
arXiv:astro-ph/0604069.

\bibitem{Chen:2006xj}
X.~Chen, R.~Easther and E.~A.~Lim,
JCAP {\bf 0706}, 023 (2007).

\bibitem{Zehavi:1998gz}
I.~Zehavi, A.~G.~Riess, R.~P.~Kirshner and A.~Dekel,
Astrophys.\ J.\  {\bf 503}, 483 (1998).

\bibitem{Jha:2006fm}
S.~Jha, A.~G.~Riess and R.~P.~Kirshner,
Astrophys.\ J.\  {\bf 659}, 122 (2007).

\bibitem{Conley:2007ng}
A.~Conley, R.~G.~Carlberg, J.~Guy, D.~A.~Howell, S.~Jha, A.~G.~Riess
and M.~Sullivan,
arXiv:0705.0367 [astro-ph].

\bibitem{Drexlin:2004as}
G.~Drexlin,
Eur.\ Phys.\ J.\  C {\bf 33}, S808 (2004).

\bibitem{Lewis:2002ah}
A.~Lewis and S.~Bridle,
Phys.\ Rev.\  D {\bf 66}, 103511 (2002).

\bibitem{gelman}
A.~Gelman and D.~B.~Rubin, 
Stat. Sci. {\bf 7}, 457 (1992).

\bibitem{Dunkley:2004sv}
J.~Dunkley, M.~Bucher, P.~G.~Ferreira, K.~Moodley and C.~Skordis,
Mon.\ Not.\ Roy.\ Astron.\ Soc.\  {\bf 356}, 925 (2005).

\end{thebibliography}
\end{document}